\documentclass[12pt]{article}
\pdfoutput=1
\usepackage{jheppub}
\usepackage{simplewick}
\usepackage{verbatim}
\usepackage{graphics, color}
\usepackage{graphicx}
\usepackage{amsmath}
\usepackage{amssymb}
\usepackage{amsthm}
\usepackage{amsfonts}
\usepackage[usenames,dvipsnames]{xcolor}
\usepackage[normalem]{ulem}

\newcommand{\be}{\begin{equation}}
\newcommand{\ee}{\end{equation}}

\newcommand{\tr}{\mathrm{tr}}

\newcommand{\lan}{\langle}
\newcommand{\ran}{\rangle}
\newcommand{\Tr}{\mathrm{Tr}}

\newcommand{\mO}{\mathcal{O}}

\newcommand{\A}{\mathcal{A}}

\definecolor{grey}{rgb}{.5,.5,.5}
\definecolor{bluegreen}{rgb}{0,.5,.5}
\definecolor{darkgreen}{rgb}{0,.5,0}

\newcommand{\AAl}[1]{{\textbf{\textcolor{red}{#1}}}}

\newcommand{\BS}[1]{{\textbf{\textcolor{blue}{#1}}}}

\newcommand{\beq}{\begin{equation}}
\newcommand{\eeq}{\end{equation}}
\def\({\left(}
\def\){\right)}

\begin{document}

\title{Linearity of Holographic Entanglement Entropy}
\author{Ahmed Almheiri,$^a$}
\author{Xi Dong$^b$ and}
\author{Brian Swingle$^a$}
\affiliation[a]{Stanford Institute for Theoretical Physics, Department of Physics, Stanford University, Stanford, CA 94305, USA}
\affiliation[b]{School of Natural Sciences, Institute for Advanced Study, Princeton, NJ 08540, USA}
\emailAdd{almheiri@stanford.edu}
\emailAdd{xidong@ias.edu}
\emailAdd{bswingle@stanford.edu}
\abstract{We consider the question of whether the leading contribution to the entanglement entropy in holographic CFTs is truly given by the expectation value of a linear operator as is suggested by the Ryu-Takayanagi formula. We investigate this property by computing the entanglement entropy, via the replica trick, in states dual to superpositions of macroscopically distinct geometries and find it consistent with evaluating the expectation value of the area operator within such states. However, we find that this fails once the number of semi-classical states in the superposition grows exponentially in the central charge of the CFT. Moreover, in certain such scenarios we find that the choice of surface on which to evaluate the area operator depends on the density matrix of the entire CFT. This nonlinearity is enforced in the bulk via the homology prescription of Ryu-Takayanagi. We thus conclude that the homology constraint is not a linear property in the CFT. We also discuss the existence of ‘entropy operators’ in general systems with a large number of degrees of freedom.}

\maketitle
\addtocontents{toc}{\protect\enlargethispage{2\baselineskip}}

\section{Introduction}
Entropy is not a linear operator while area is, yet in gravity these two quantities are usually equated.

This was first observed in the context of black hole thermodynamics where it was shown that the entropy of a black hole is given by the expectation value of the area operator evaluated on its event horizon \citep{Bekenstein:1973ur}. This operator is a nonlinear functional of the canonical variables of quantum gravity (the metric and conjugate momentum) and is understood to be a linear operator which maps states to states. Given that this entropy is a coarse-grained thermodynamic quantity, it seems plausible that it can be represented by a linear operator very much in the same way that the entropy of a gas can be represented by its energy. One should probably expect this property in systems with a thermodynamic limit and which are known to thermalize.

A more paradoxical relationship between entropy and area arises in the context of the AdS/CFT correspondence. This correspondence is a duality between string theories living in $d+1$-dimensional  asymptotically Anti de Sitter (AdS) space and certain $d$-dimensional conformal field theories (CFTs) which can be thought of as living on the boundary of AdS \citep{Maldacena:1997re}. One way in which these two descriptions are connected is via the identification of the central charge of the CFT with the ratio of  the AdS length to the Planck length to some positive power,  $c \sim (L_{AdS}/l_P)^\# $. This duality provides a nonperturbative definition of a certain class of theories of quantum gravity in asymptotically AdS spacetimes in terms of a certain class of CFTs.

An outcome of this duality is that the strong coupling and $c \rightarrow \infty$ limit of the CFT is described on the AdS (bulk) side by classical gravity with a gravitational constant $G_N \sim 1/c^\#$, demonstrating the strong/weak dual nature of AdS/CFT. It is in this limit that a remarkably simple, albeit confusing, formula for the entanglement entropy of any region of the CFT was proposed \citep{Ryu:2006bv}. It was suggested that, in static situations, the entanglement entropy of a subregion $R$ of the CFT  is given by the area of the minimal area bulk surface $X$ anchored to the boundary of $R$, $\partial X = \partial R$, and homologous to $R$, denoted by $X \overset{h}{\sim} R$,
\begin{align}
S_R = {A(X_{min}) \over 4 G} \Big|_{\partial X = \partial R}^{X \overset{h}{\sim} R}. \label{RT}
\end{align}
We shall refer to this henceforth as the RT formula. This formula was proven in \citep{Lewkowycz:2013nqa} under certain reasonable assumptions including the extension of the replica symmetry into the dominant bulk solution. It was also extended to the time dependent case in \citep{Hubeny:2007xt} where the minimal surface generalizes to a spacelike extremal surface. Another proposal for the time dependent case was presented in \citep{Wall:2012uf}; their prescription was to find the minimum area $X$ on every possible spatial slice containing the interval $R$, and then to pick out from this set the one with maximal area. Our focus here will be mostly on the static case.

In the same way that the Bekenstein-Hawking entropy receives corrections from entanglement of quantum fields across the horizon \citep{Bekenstein:1973ur, Hawking:1971tu}, the entanglement entropy of a region of the CFT also gets corrected \citep{Faulkner:2013ana}. This analogy was spelled out more generally in \citep{Engelhardt:2014gca} which discusses further corrections to the entanglement entropy. However, all these corrections are subleading in $c$, and are manifestly not given by the expectation value of linear operators. At leading order in $c$ there will also be higher derivative corrections involving various curvature invariants which are linear operators like the area. Our focus in this paper is solely on the linear nature of the leading area term of the entanglement  entropy, and so will center the discussion mainly on the RT formula.

In contrast to the notion of entropy in black hole thermodynamics, formula \ref{RT} equates the expectation value of the area operator with a truly microscopic measure of information. This microscopic measure, or entanglement entropy, is given by
\begin{align}
S_R(|\psi \rangle) = - tr_{R} \rho_R \ln \rho_R = \langle \psi |\left( - \ln \rho_R \right)| \psi \rangle,
\end{align}
where $-\ln \rho_R$ is an operator on $R$ alone, and  returns the correct entropy only for the specific state $|\psi\rangle$. One can try to extend this definition to apply to a basis of states $| \psi_i \rangle$ and construct the following entropy operator,
\begin{align}
\hat{S}_R = -\sum_i P_i \ln\left(  Tr_{\bar{R}} | \psi_i \ran \lan \psi_i |   \right) P_i.
\end{align}
This operator would produce the correct result for any element of the chosen basis. However, it will it general fail to do so for linear combinations. Take for example the two-qubit Hilbert space spanned by the product states $|i j \ran$, with $i, j \in \{0,1 \}$. The entropy operator for a single qubit should have zero expectation value for any state in this basis. Since this statement is also true for any other product basis, we conclude (erroneously) that the entropy operator is zero.

The confusing aspect of the RT formula is that it seems makes the replacement
\begin{align}
\left( - \ln \rho_R \right) \rightarrow {\hat{A}(X_{min}) \over 4 G} \Big|_{\partial X = \partial R}^{X \overset{h}{\sim} R},
\end{align}
thus, identifying the area operator with the modular Hamiltonian \cite{Jafferis:2014lza, Jafferis:2015del}. Just as $ - \ln \rho_R$ was state dependent, the surface on which the area operator is evaluated, $X_{min}$, depends on the dual geometry and consequently on the state. One might then be tempted to generalize the RT proposal to
\begin{align}
&S_R(|\psi\rangle) = \langle \psi | {\hat{\cal{A}}} | \psi \rangle, \label{RToperator} \\
&\hat{\cal A} \equiv \sum_{i} {A(X^i_{min}) \over 4 G} P_i, \label{areaopno}
\end{align}
where the projection operators, $P_i$, project onto subspaces of states with the same classical geometry, and $X^i_{min}$ is the extremal surface in that geometry. We also removed the operator symbol from the area term due to the presence of the projection operators. Moreover, this construction assumes that we are working at leading order in the $1/c$ expansion. To this accuracy, different semi-classical states are orthogonal and this operator is block diagonal in this basis allowing for no off-diagonal terms between states of different geometries.

However, as we will argue below, a minimal area operator can be constructed as a gauge invariant linear operator in the Hilbert space. Thus, it is sufficient to generalize RT by simply writing the area operator as
\begin{align}
\hat{\cal A} \equiv  {\hat{A}(X_{min}) \over 4 G}  \label{areaopyes}
\end{align}
where now $X_{min}$ will be operator valued and will specify the location of the minimal area surface in any geometry. We will investigate how we expect the off-diagonal elements of this operator to behave.

The goal of this paper is to study the applicability of this interpretation of RT beyond semi-classical states. Since the question in focus is about linearity, we investigate what both sides of equation \ref{RToperator} produce for states dual to macroscopic superpositions of distinct bulk geometries. We are thus considering an `extended RT proposal' which asserts that the entanglement entropy of a subregion in the CFT is still given by the expectation value of the area operator within states dual to superpositions of geometric states. This extended RT proposal does not follow trivially from the RT formula (or its derivations) for a single geometry; we will nevertheless provide evidence in favor of the extended RT proposal within certain limits. Since it is crucial to compare the calculation on both sides of the duality, we will focus on the context of AdS$_3/$CFT$_2$, relying heavily on computational techniques of holographic 1+1 CFTs. Our main probe will be the entanglement entropy of a single interval on the cylinder. What we will find is that the entropy behaves like a linear operator within a large class of subspaces of semi-classical states of dimension less than $e^{O(c)}$.

Is this result general? We argue yes. By analogy with thermodynamics, where changes in entropy are related to fluxes of energy (a manifest observable), our proposal is that a thermodynamic or large $N$ limit is sufficient to have entropies which behave as the approximate expectation value of a linear operator. Our result that entropies average in two dimensional holographic CFTs supports this proposal. We also exhibit an information theoretic setting where entropy behaves as the expectation of a linear operator. The key idea is that in an appropriate thermodynamic limit the entropy can be determined by performing a measurement which only weakly disturbs the state. Finally, we discuss a number of related issues including the non-linearity of the Renyi entropy, the precise limits of linearity, and the role of strong coupling.

A similar proposal has been sketched by \cite{Papadodimas:2015jra}; we discuss in more detail the relationship between their proposal and our work in the discussion.

\section{The Area Operator of the Ryu-Takayanagi Proposal}

In this section, we define the quantities that appear on the right hand side of equation \ref{RToperator} in some more detail. Firstly, this formula was  checked in $1+1$ holographic CFTs for many states dual to semi-classical geometries\footnote{We do not exclude states with non-classical bulk regions provided those bulk regions are not probed by the Ryu-Takayanagi surface associated with any boundary region. These states might be characterized as having an extremal surface barrier \citep{Engelhardt:2013tra} shielding the non-classical regions.}, i.e. small quantum fluctuations on a fixed gravitational background, where the notion of area is unambiguously well defined. The natural interpretation of these states is as coherent states constructed from the metric and conjugate momenta and are highly peaked about some classical solution of Einstein's equations. The fluctuations about the classical solution are suppressed by a power of $\hbar$ which is controlled by some negative power of the central charge, $c$.

\subsection{A Gauge Invariant Area Operator}
The area operator  $\hat{A}(X_{min})$  is an operator in quantum gravity and needs to be defined in a gauge invariant way \citep{DeWitt:1967yk}. This is usually ensured by defining the operator with respect to something fixed under gauge transformations \citep{Giddings:2005id}. In the specific case of AdS, pure gauge diffeomorphisms  are those that keep the boundary conditions of AdS fixed \citep{Marolf:2015jha, Donnelly:2015hta}. Thus,  $\hat{A}(X_{min})$ needs to be completely specified by boundary data to be gauge invariant. In particular, the curve $X_{min}$ needs to be localized in a gauge invariant way, i.e. determinable purely from some boundary data.

The bulk interpretation of the area operator \ref{areaopyes} on a state can be achieved with the following prescription. Starting with a $| \psi \ran$ of the entire CFT, one can, in principle, construct the background geometry with something like the HKLL formalism \citep{Hamilton:2006az, Heemskerk:2012np, Kabat:2012hp} as expectation values of geometric bulk fields. We can then consider all co-dimension 2 surfaces on this geometry that are anchored on the boundary of $R$ and homologous to it and find the specific one that extremizes the area. If there are multiple such surfaces we simply take the one with smallest area. This locates $X_{min}$ in a gauge invariant way. Since the area operator is then evaluated on this surface, it is also gauge invariant.

\subsection{The Boundary Support of the Area Operator}
\label{entwedgerecon}
The next thing to determine is the support of $\hat{A}(X_{min})$ in the CFT. In the bulk, this operator lies on the edge of the entanglement wedge of the region $R$. The entanglement wedge is defined as the domain of dependence of a spacelike surface bounded by $R$ and $X_{min}$ \cite{ent_wedge_def}. The part of the bulk this process carves out is the entanglement wedge. That all operators within the entanglement wedge have representations only on $R$ has been argued for in \citep{Almheiri:2014lwa}, and proven recently in \citep{Dong:2016eik}. We will take this point of view for the rest of this paper. One might worry about $\hat{A}(X_{min})$ acting on the edge of the entanglement wedge and whether that really should be considered as part of the wedge. This can be dealt with by defining $\hat{A}(X_{min})$ in a limit sense; follow the same prescription for a slightly smaller interval, that area operator is guaranteed to lie within the entanglement wedge of $R$ \citep{Engelhardt:2013tra}, and then take the limit as the intervals become the same size.

\subsection{The Linearity of the Area Operator}
A commonly raised question about the minimal area operator of the RT proposal is whether it is state dependent in the same way as the entanglement entropy. We argue that while it is certainly state dependent with regards to picking a different surface for each state it nevertheless is still a linear operator. This is a mild form of state dependence, otherwise known as background dependence \citep{Heemskerk:2012mn}, unlike what is found with the entanglement entropy.

We described in the introduction, around equation \ref{areaopno}, how the minimal area operator can be constructed to leading order in the $1/c$ expansion by using projection operators that project onto subspaces with the same background geometry. This operator is by construction a linear operator and is block diagonal in a semi-classical basis. Thus, it contains no off-diagonal terms between states of different background geometry.

Here we want to present a different definition of the minimal area operator which allows for the presence of off-diagonal terms. We make no statement about the uniqueness of this construction but believe that all definitions will behave more or less in the same way. In particular, they should all agree with \ref{areaopno} in the infinite $c$ limit.

First, let us write the area operator that is evaluated on some surface $\hat{X}$. We will shortly specify more carefully how $\hat{X}$ is defined. The area is
\begin{align}
\hat{A}(\hat{g}, \hat{X})  = \int d\hat{y} \sqrt{ \hat{g}(\hat{y})}. \label{linearArea}
\end{align}
The measure $d\hat{y}$ should be understood as determining the domain of the integral as localized on the surface $\hat{X}$. In order for this quantity to be well defined, we need to specify $\hat{X}$ in a gauge invariant way. This can be achieved by pinning down its location relationally in terms of proper distances to some boundary points $b$ via an operator relation $d_{\hat{g}} \big[ y,b; \hat{\theta}(b) \big] = \hat{f}(b)$; the metric dependence comes in from the definition of the proper distance between $\hat{y}$ and $b$. The operator $\hat{\theta}(b)$ specifies along which geodesic to travel into the bulk and the operator $\hat{f}(b)$ determines the amount of proper distance required to reach the surface. Together, they determine the shape and location of the bulk surface and will be determined shortly by the minimization condition. Inverting this relation gives
\begin{align}
\hat{y}(b) = d^{-1}_{\hat{g}}\left[ \hat{f}(b), b, \hat{\theta}(b) \right].
\end{align}
This process is only consistent if $\hat{y}$ is an operator since it depends on the metric field operator, $\hat{f}$, and $\hat{\theta}$. Plugging this back into \ref{linearArea} we obtain
\begin{align}
\hat{A}(\hat{g}, \hat{f}, \hat{\theta}) &= \int db \ {\cal J}^{\hat{g}}_{ \hat{f}(b), \hat{\theta}(b)} \sqrt{- \hat{g}\left(d^{-1}_{\hat{g}}\left[ \hat{f}(b), b, \hat{\theta}(b)\right]\right)}, \label{areag}
\end{align}
where ${\cal J}^{\hat{g}}_{ \hat{f}(b), \hat{\theta}(b)}$ is a Jacobian factor which depends on $\hat{g}$, $\hat{f}$, and $\hat{\theta}$. Thus, we can think of the area operator as a function of the background metric $\hat{g}$ and of $\hat{f}$ and $\hat{\theta}$ which specify the surface. To get the minimal area operator we simply require that it is minimal with respect to $\hat{f}$ and $\hat{\theta}$,
\begin{align}
& \delta_{\hat{f}} \hat{A}(\hat{g}, \hat{f} , \hat{\theta}) = 0 \\
& \delta_{\hat{\theta}} \hat{A}(\hat{g}, \hat{f}, \hat{\theta} ) = 0
\end{align}
as an operator equation. The next step would be to solve for $\hat{f}$ and $\hat{\theta}$ in terms of $\hat{g}$ and plug it back into \ref{areag}. This procedure would ultimately produce a minimal area operator as a function purely of the metric.

To be clear, the construction above is formal and, for example, involves non-polynomial functions of the metric. Although it is beyond the scope of our work, it is possible that there is a fully non-perturbative definition of the bulk quantum gravity in which case the formal manipulations above might yield a non-perturbatively defined linear area operator. Alternatively, we may work perturbatively around semi-classical solutions in which case the above formal manipulations can be used to define an operator order-by-order in perturbation theory. Such a perturbative construction is sufficient for most of our statements and amounts to working with a space of states defined on top of the quantum state describing the classical solution. In the context of AdS/CFT, an intriguing possibility is that there exists a CFT operator defined on the whole Hilbert space which approximately reduces to the perturbatively defined area operator around any given saddle.

\subsection{The Area Operator on Superpositions -- A Prediction}
\label{subsec:areaop_predict}
Having shown that the RT area operator computes a physical gauge invariant quantity in the bulk, it is plausible to assume that it is given by a linear operator when acting on any subspace spanned by semi-classical states. The goal of this subsection is to understand the structure of the off-diagonal components of the area operator within such a subspace.

The correct way to think of the area term appearing on the right hand side of RT is as the saddle point evaluation of the expectation value of the area operator,
\begin{align}
\big{\langle} \hat{A}\left( \hat{g} \right) \big{\rangle} \approx A(g_{s}),
\end{align}
where $g_s$ is the dominant saddle point of the partition function. Note that this is an ${\cal O}(c^0)$ number; here we are evaluating the expectation value of the area operator without the factor of $1/G_N$. One way to see this is via the generating function of moments of $A$,
\begin{align}
Z(J) = \int \mathcal{D} g \, e^{J A(g)} e^{- c S(g)}.
\end{align}
Because $A$ involves no explicit $c$ dependence, the dominant saddle point of $Z[J]$ approaches the dominant saddle point of $Z[0]$ for any fixed $J$ as $c \rightarrow \infty$. This implies that the fluctuations of $A$ go to zero,
\begin{align}
\big{\langle} \big[\hat{A}\left( \hat{g} \right) \big]^2 \big{\rangle} - \big{\langle} \hat{A}\left( \hat{g} \right) \big{\rangle}^2 \approx 0 \ \mathrm{as} \ G_N \sim 1/c \rightarrow 0.
\end{align}
Thus, semi-classical states that can be prepared using the path integral become eigenstates of the area operator in the infinite $c$ limit.\footnote{The same result can be obtained by taking $c \rightarrow \infty$ with $J/c$ fixed, differentiating with respect to $J/c$, and then taking $J/c \rightarrow 0$. If we differentiate but do not set $J/c \rightarrow 0$, then we are computing the flucutation of the area operator around a different saddle point. These two ways of computing the flucutation around the original saddle point will agree provided the limits $c\rightarrow \infty$ and $J \rightarrow 0$ commute.} The rest of this section will investigate the nature of the suppression of the off-diagonal elements of the area operator.

\vspace{3mm}
\noindent\textbf{States with energy $O(c^0)$}

Consider first the subspace of states of energy  $O(c^0)$. All of these states are dual to pure AdS with a very diffuse gas of particles or possibly black holes whose mass does not scale with $c$. Einstein's equations predict that the deformation of this stress energy away from pure AdS will be suppressed by $1/c$. To see this, consider the linearized form of Einstein's equations
\begin{align}
\Box^{\mu \nu}_{\alpha \beta} h_{\mu \nu} = G_N T_{\alpha \beta}, \label{lingr}
\end{align}
where $h_{\mu \nu}$ is a perturbation of the metric, $T_{\alpha \beta}$ is the stress energy of matter in AdS, and $\Box$ is some differential operator. The background spacetime is determined by the sourceless Einstein equations. The deformation of the area of a surface away from the background value is controlled by $h$ as
\begin{align}
\delta A & \sim \int_{\Sigma}   \ (\dots)^{\mu \nu} \ h_{\mu \nu} \\
& \sim G_N \int_{\Sigma}   \ (\dots)^{\mu \nu} \int K_{\mu\nu}^{\alpha \beta} \ T_{\alpha \beta} \label{defa}
\end{align}
where $K$ is the Green's function solving \ref{lingr}.

To estimate the cross terms, we first promote equations \ref{lingr} - \ref{defa} to operator equations. Then the area operator will have the form
\begin{align}
\hat{A}  = \int_{\Sigma}  \sqrt{-g} \ +  G_N \int_{\Sigma}   \ (\dots)^{\mu \nu} \int K_{\mu\nu}^{\alpha \beta} \ \hat{T}_{\alpha \beta}, \label{areastressop}
\end{align}
and we can directly compute its matrix elements. These will be
\begin{align}
\lan i | \hat{A} | j \ran = \delta_{i j} \ \int_{\Sigma}  \sqrt{-g} \ +  G_N \int_{\Sigma}\  (\dots)^{\mu \nu}  \int K_{\mu\nu}^{\alpha \beta} \lan i | \hat{T}_{\alpha \beta} | j \ran.
\end{align}
Since the matrix elements of the stress tensor is $O(c^0)$ within this subspace, we conclude that the off-diagonal elements in this subspace is suppressed by $G_N \sim 1/c$. Notice also that the eigenvalues degenerate in this limit.

If we consider an arbitrary state within this low energy subspace, one might worry that the small off-diagonal terms could potentially add up and compete with the diagonal terms. However, due to the sparseness condition on the CFT, the dimension of this subspace is not large enough to ever make the off-diagonal terms matter; the number of states of energy $O(c^0)$ is $\ll O(c)$ and so the off-diagonal contribution will always be $O(c^{-1})$. Thus we conclude that the area of the minimal surface for any state in this subspace is same to leading order in $c$.

\vspace{3mm}
\noindent\textbf{States with energy scaling with $c$}

At energies scaling with the central charge, it is characteristic of holographic theories to have a fairly dense spectrum possibly admitting a statistical description. Here we have in mind using the \emph{eigenstate thermalization hypothesis} (ETH) \citep{PhysRevA.43.2046, 0305-4470-29-4-003} to conjecture a form for the area operator at high energies. ETH states that the expectation value of a suitably coarse operator in an energy eigenstate is given by its microcanonical average. This statement is supposed to hold for states with a finite energy density, meaning
\begin{align}
\lim_{c\rightarrow \infty} \frac{E - E_g}{c} > 0
\end{align}
where $E_g$ is the ground state energy. Since the notion of geometry is expected to be an emergent coarse phenomenon of holographic theories, one would expect that the spectrum of the operator which probes this geometry to be dictated by ETH. Assuming ETH, the form of the area operator in an energy eigenstate basis at high energies will be
\begin{align}
\hat{A}_{\alpha \beta} = A(E_{\alpha \beta}) \delta_{\alpha \beta} + e^{- S(E_{\alpha \beta})/2} f(E_{\alpha \beta}) R_{\alpha \beta} \label{areaeth}
\end{align}
where $E_{\alpha \beta} = {E_\alpha + E_\beta \over 2}$ and $A$, $f$ are smooth functions of the average energy. $R_{\alpha \beta}$ is an erratic function of $\alpha$ and $\beta$ with zero mean and unit average magnitude. $S(E_{\alpha \beta})$ is the logarithm of the number of states between $E_\alpha$ and $E_\beta$.

To get a sense of the structure of energy eigenstates, we follow Hawking and Page \citep{Hawking:1982dh} and consider a system composed of thermal gas and black holes in more than three dimensions. Let us focus on a microcanonical ensemble of states centered around an energy $E$ with width of order $c^0$. The dominant state within this ensemble can be determined by comparing the number of states, or the entropy, of the possible configurations with energy $\sim E$. In comparing a thermal gas of light particles in AdS and a black hole, one finds four possible phases. Below some energy $E_0$ all black holes evaporate and the dominant state is a thermal gas. Above a higher energy $E_2$ all configurations of gas collapse to form a black hole. Between these two energies there exists stable configurations of either a gas or a small black hole, but which configuration dominates depends on the energy. Across some energy $E_1$ within this window the dominance of the two configurations switches from gas to black hole as the energy is increased. The restriction to greater than three dimensions arises because there are no small black holes in AdS$_3$, but in more complicated examples coming from string theory the phase structure can be much richer and can include small ``enigmatic" black holes \cite{deBoer:2008fk, Bena:2011zw}.

However, we must be cautious in applying ETH reasoning to microcanonical phases with both gas and black hole states because the energies involved scale like $c^a$ with $a<1$ so these states do not lie within the traditional regime of validity of ETH. Very large black holes, with energy scaling like $c$, always have $E > E_2$ and hence reside in a regime where the only stable solutions are black holes. For such energies it is plausible that all microstates ``look the same" geometrically and have small off diagonal matrix elements for the area operator in accord with ETH. Assuming also that the area operator is a coarse operator then the off-diagonal matrix elements of the area operator can be neglected until we consider an exponentially large superposition of microstates.

For states of intermediate energy we cannot make as strong a statement. We would expect matrix elements of the area operator between different energy eigenstates to be at least of order $O(c^{-1})$, so that one can still superpose a small number of microstates while neglecting off diagonal matrix elements. It is not even clear if the microstates are geometric at intermediate energies, say between $E_0$ and $E_2$. It is possible that ETH could still apply with a different notion of energy density, i.e. keeping $\frac{E-E_0}{c^a} > 0 $ as $c \rightarrow \infty$.

It might also be possible to construct sets of wave packets, each consisting of many microstates, such that the corresponding states are approximately stationary (on shorter than exponential times) and are approximately geometrical, being either approximately a black hole or approximately a thermal gas. Within sets of such approximate black hole states, say, we might again suspect that the matrix elements of the area operator are exponentially small.

\vspace{3mm}
\noindent\textbf{Summary and prediction}

We have presented plausible reasons for thinking that the area operator behaves as a coarse operator and should have suppressed off-diagonal components. We discussed how the area operator maintains the same result for any state within a low energy subspace of energy $O(c^0)$. At higher energies something different happens. Consider the expectation value of the area operator in an arbitrary state
\begin{align}
| \psi \ran = \sum_\alpha^M c_\alpha | E_\alpha \ran
\end{align}
within a small shell of high energy way above the Hawking Page transition. Using the ETH form of the area operator \ref{areaeth} this is
\begin{align}
\lan \psi | \hat{A} | \psi \ran = A(E) + e^{- S(E)/2} f(E) \sum_{\alpha \beta}^M c^*_\alpha R_{\alpha \beta} c_\beta \label{psiareapsi}
\end{align}
where we have assumed that the functions $A$, $S$, and $f$ are more or less constant within the considered energy window. Recall that the matrix $R_{\alpha \beta}$ oscillates wildly as a function of its indices. Thus, for an arbitrary state with random $c_\alpha$'s the sum above will be highly suppressed. In fact, even if we pick all the $c_\alpha$'s to be equal, it would still not contribute. The only way to deviate from the microcanonical average is by carefully choosing the coefficients to correlate with the fluctuations in $R_{\alpha \beta}$. Even with this fine tuning, this sum can at most be $M^{1/2}$ \footnote{$c_\alpha R_{\alpha \beta} c_\beta$ is just the largest eigenvalue of matrix $R$. If $R$ is a random $M \times M$  Hermitian matrix with all matrix elements independent and normally distributed, then the maximum eigenvalue is known to be of order $M^{1/2}$ \cite{tracywidom1,tracywidom2}. In our case, $M \sim e^{S(E)}$ so the largest eigenvalue is of order $e^{S(E)/2}$.}  In order to deviate by an order one amount from the microcanonical average, the state must consist of a finely tuned superposition involving $e^{S(E)}$ states.

The result of this subsection is that as long as we don't consider finely tuned states of $e^{O(c)}$ terms then the expectation value of the area operator will simply be the average of the area in each branch of the wavefunction. Therefore, we can combine this with the RT proposal and make a prediction for the behavior of entropy within such a superposition. In particular we predict that
\begin{align}
\lim_{c\rightarrow \infty} \frac{S_R\Big(\sum_i \alpha_i | \psi_i \ran \Big) - \sum_i |\alpha_i|^2 S_R\Big(| \psi_i \ran\Big)}{c} = 0 \label{pred}
\end{align}
for a superposition of semiclassical states $| \psi_i \ran$. We will confirm this prediction in the following sections to come.

\section{How to Compute Entanglement Entropy in $1+1$ CFTs}
\label{review}

Let us review how entanglement entropy of subregions is computed in 1+1 CFTs. We describe the procedure for arbitrary subregions in general states and discuss the simplifications which occur in holographic CFTs. We will explicitly perform the calculation for a single interval in a primary state. This will mostly be a summary of \citep{Calabrese:2004eu, Asplund:2014coa}.

\subsection{Entanglement Entropy and the Replica Trick}
The entanglement entropy, also called the von Neumann entropy, of a subsystem $R$ of a quantum system is given by
\begin{align}
S_R = - tr \rho_R \ln \rho_R,
\end{align}
where $\rho_R$ is the density matrix of $R$ obtained by tracing over the rest of the system,  $\rho_R = tr_{\bar{R}} | \psi \ran \lan \psi |$. This quantity is usually technically difficult to compute in a quantum field theory due to the logarithm, but can be simplified by using the so-called `replica trick' to re-express it as
\begin{align}
S_R = \lim_{n \rightarrow 1} {1 \over 1-n}\ln \left[ tr \rho_R^n \right] \equiv \lim_{n \rightarrow 1} S^R_n,
\end{align}
where $S^R_n$ is called the $n^{th}$ Renyi entropy of $R$. Since the trace of any density matrix is one and all its eigenvalues are positive definite, one can show that the Renyi entropies are absolutely convergent and analytic for all $\mathrm{Re}[ n] >1$ \footnote{This statement is true for any quantum system of finite total Hilbert space dimension, for example, a quantum field theory in a finite box with a lattice regulator. New singularities can appear when the number of degrees of freedom goes to infinity.}. This justifies  the continuation of $n$ and allows one  to represent the entropy as
\begin{align}
S_R = \lim_{n \rightarrow 1} S^R_n  = - \lim_{n \rightarrow 1^+} \partial_n tr \rho_R^n. \label{entdrho}
\end{align}

Thus, the problem of finding the entropy has been reduced to computing the trace of the $n^{\mathrm{th}}$ power of the density matrix as an analytic function of $n$. This latter task can be implemented by evaluating the partition function of the theory on the replicated manifold $\mathbb{C}^n$ with the different sheets identified across the interval $R$ \citep{Calabrese:2004eu}. With the appropriate normalization this is
\begin{align}
tr \rho_R^n = {Z_n \over Z_1^n},
\end{align}
where $Z_1$ is the partition function of the CFT in question. When computing the entropy in an arbitrary state $| \psi \ran$, $Z_1$ is given by $\lan \psi  | \psi \ran$. $Z_n$ is the `replicated' partition function obtained by gluing  $n$ copies of the original CFT along the region $R$. Note that the replicated density matrix satisfies the condition that $tr \rho_R^n \rightarrow 1$ as $n \rightarrow 1$.

It turns out there is a further simplification for computing this quantity. Let us consider the case where  $R$ is a subregion  composed of $N$ disjoint intervals. By considering the expectation value of the stress tensor within the replicated partition function \citep{Calabrese:2004eu}, one can show that that $Z_n$ can be written as the 2$N$-point function of so-called twist operators,
\begin{align}
tr \rho_R^n = {Z_n \over Z_1^n} = \lan \psi^{\otimes n}| \prod_i^N \sigma_n(u_i, \bar{u}_i) \sigma_{-n} (v_i, \bar{v}_i) | \psi^{\otimes n} \ran, \label{rhon}
\end{align}
where this expectation value is evaluated in the orbifold theory on $\mathbb{C}^n$ and $|\psi^{\otimes n}\rangle = \bigotimes_{i=1}^n|\psi \rangle $. The coordinates $u_i$ and $v_i$ are the endpoints of the intervals. In this theory, the twist operators behave as primary operators of  dimensions
\begin{align}
h_{twist} = \bar{h}_{twist} =  {c \over 12} \left( n - {1 \over n} \right)
\end{align}
and vanishing spin.

\subsection{A Single Interval Example}

\label{example}
\begin{figure}
\begin{center}
\includegraphics[height=6cm]{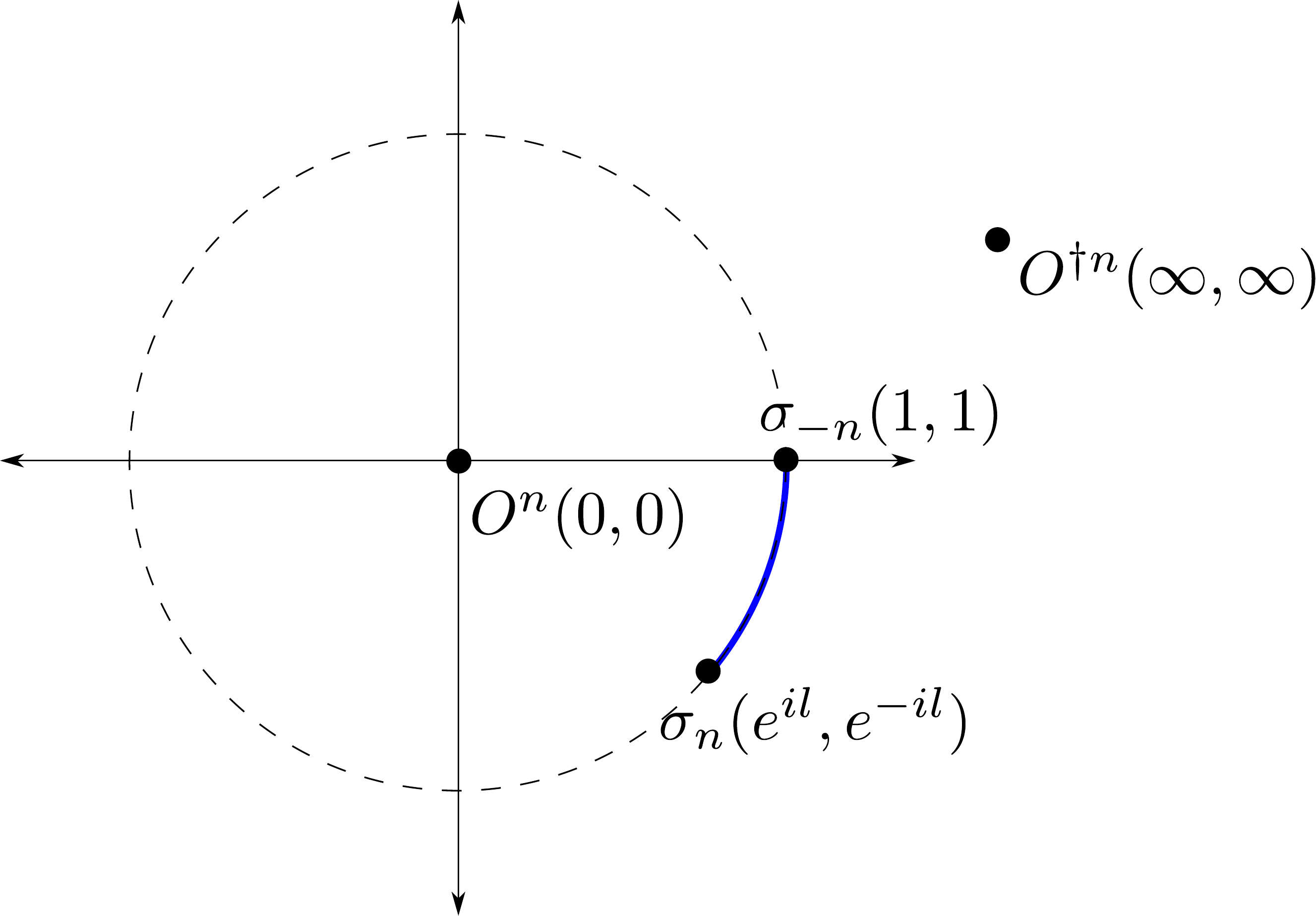}
\caption{The configuration of operators in the four-point function expression of the replicated density matrix. The blue line represents the subregion $R$ of the CFT. The twist operators are restricted to the unit circle representing a single spatial slice of the cylinder.}\label{otto}
\end{center}
\end{figure}

Now we specialize to computing the entanglement entropy of a single interval on the cylinder in an excited state. We will consider an arbitrary primary state prepared in the usual way using the state-operator correspondence,
\begin{align}
| \mO \rangle \equiv \lim_{x,\bar{x} \rightarrow 0} \mO(x,\bar{x}) | 0 \rangle,
\end{align}
where $\mO$ is an arbitrary primary operator of dimensions $h, \ \bar{h}$. The conjugate of this state is defined as
\begin{align}
\langle \mO | \equiv \lim_{x, \bar{x} \rightarrow \infty} x^{2 h} \bar{x}^{2 \bar{h}} \langle 0 | \mO^{\dagger}(x,\bar{x}),
\end{align}
which ensures the state is normalized to one. The trace of the replicated density matrix on an interval $R$ in this state is
\begin{align}
tr \rho^n_R &=  \lan \mO^{\otimes_n} |  \sigma_n(z, \bar{z}) \sigma_{-n} (1, 1) | \mO^{\otimes_n} \ran \\
&=  \lan 0^{\otimes n} |  \underbrace{\mO^{\dagger}\otimes  ... \otimes\mO^{\dagger}}_{n}   \sigma_n(z, \bar{z}) \sigma_{-n} (1, 1)  \underbrace{\mO \otimes ... \otimes  \mO}_{n}| 0^{\otimes n}  \ran. \label{replden}
\end{align}
The location of $z$ and $\bar{z}$ will be restricted to the unit circle on the $x$-plane; this chooses a preferred, and natural,  time slicing of the CFT on the cylinder. The locations of these operators is presented in figure \ref{otto}.

This is a four point function of primary operators in the orbifold theory on $\mathbb{C}^n$. We can use the techniques of conformal blocks to compute this expression. By performing an operator product expansion (OPE) in the $t$-channel of  the two tensor product operators together and the two twist operators together we get
\begin{align}
\lan 0 |  ( \mO^{\dagger} )^n  \sigma_n(z, \bar{z}) \sigma_{-n} (1, 1) \left(   \mO \right)^n| 0  \ran = \sum_p C^{O^n O^n}_p C^{\sigma_n \sigma_n}_p {\cal F}^{O^n O^n}_{\sigma_n \sigma_n} \left(p |  1 - z \right) \bar{{\cal F}}^{O^n O^n}_{\sigma_n \sigma_n} \left( p | 1 - \bar{z} \right), \label{4pnt}
\end{align}
where the sum $p$ is over all the primary operators of the theory. Conformal invariance fixes the contribution from all the descendent operators, which are implicitly resummed to give the functions ${\cal F}$ and $\bar{{\cal F}}$. These functions are known as `conformal blocks' and are functions of the dimensions of all the operators appearing in the four point function and the internal primary operator.

We see that the entanglement entropy depends on the details of the theory through the values of the OPE coefficients $C^{i j}_k$. In holographic theories, those with large central charge and a sparse spectrum of light operators, such a four point function is dominated by the identity block contribution. The OPE coefficient of this contribution is simply $1$, giving a universal result for holographic theories. We should note, however, that this dominance of the identity block fails for states composed of many, $O(c)$, light operators as first observed in \citep{Giusto:2014aba} in the context of supersymmetric CFTs.\footnote{See also \citep{Asplund:2014coa,Giusto:2015dfa}. We thank the authors of \citep{Giusto:2014aba} for bringing this to our attention.} In this case the OPE coefficients between light operators and the highly composite operator will be proportional to the number of light operators in the composite and will scale as some positive power of $c$; one can think of this as simply the expectation value of the light operator in the state created by the composite. These non-identity contributions can then potentially compete with the identity block. We will assume in this paper that we are working with states for which the identity block dominates.

Let us specialize to the case where $\mO$ is a heavy operator of no spin, i.e. $h = \bar{h} \sim c$.  In the bulk, this dimension translates to the total mass of the spacetime, up to a factor of the AdS radius. As discussed in Section \ref{subsec:areaop_predict}, for large enough operator dimension the dominant configuration in the bulk is a black hole \citep{Hawking:1982dh}. Since the state is pure, this is more precisely a black hole microstate. The exterior of this black hole is described to a very good approximation by the standard BTZ geometry.

For $n$ greater than one, \ref{4pnt} is a four point function of heavy operators. The form of the identity block in this case is actually not known in closed form, but a perturbative expansion in $1 - z$ can be performed \citep{Hartman:2013mia}. However, a nonperturbative result can be obtained in the limit as $n \rightarrow 1$ \citep{Fitzpatrick:2014vua}. Because the dimension of the twist operators is proportional to $n-1$, the four point function in this limit becomes that of two heavy and two light operators
\begin{align}
\lan 0 |  \mO^{\dagger}  \sigma_{1 + \epsilon}(z, \bar{z}) \sigma_{-(1 + \epsilon)} (1, 1)  \mO  | 0  \ran =  {\cal F}\left(0 |  1 - z \right) \bar{{\cal F}}\left( 0 | 1 - \bar{z} \right),
\end{align}
where we took $n = 1 + \epsilon$, and restricted to the identity block term. Remember that the blocks are functions of the dimensions of the $\mO$'s and the twist operators. As discussed in \citep{Fitzpatrick:2014vua, Hartman:2013mia}, this can be obtained in closed form by solving a differential equation with nontrivial monodromy. The leading term in $\epsilon$ contribution to this four point function is
\begin{align}
\lan 0 |  \mO^{\dagger}  \sigma_{1 + \epsilon}(z, \bar{z}) \sigma_{-(1 + \epsilon)} (1, 1)  \mO  | 0  \ran =  \left[ { |z|^{1 - \alpha} | 1 - z^\alpha|^2  \over \alpha^2}  \right]^{- {c (n^2 - n)/3}} \label{renyin}
\end{align}
where $\alpha = \sqrt{1 - 24 h_i/c}$. Using eq \ref{entdrho} gives the entanglement entropy
\begin{align}
S = {c \over 3} \ln \left[ {\beta \over \pi \epsilon_{UV}} \sinh\left( {l \pi \over \beta} \right)   \right] \label{sbtz}
\end{align}
where $\beta \equiv 2 \pi/\sqrt{24 h/c - 1}$, $l$ is the size of the interval, and $\epsilon_{UV}$ is the UV cut-off. For $l < \pi$, this is precisely the answer one would get for the entanglement entropy of an interval in the thermal state given by temperature $\beta$. This is a manifestation of the fact that the geometry outside this BTZ microstate is almost identical to that in the BTZ geometry.

Naively continuing this expression to $l> \pi$ actually gives the wrong result for the entropy in that regime. In fact, since the state we considered is a rotationally symmetric pure state we should expect the entropy to be symmetric under $l \leftrightarrow 2 \pi - l$. Since the state it pure, the entropy should start to decrease once the interval encompasses more than half of the system. This is not the case for \ref{sbtz}.

\begin{figure}
\begin{center}
\includegraphics[height=6cm]{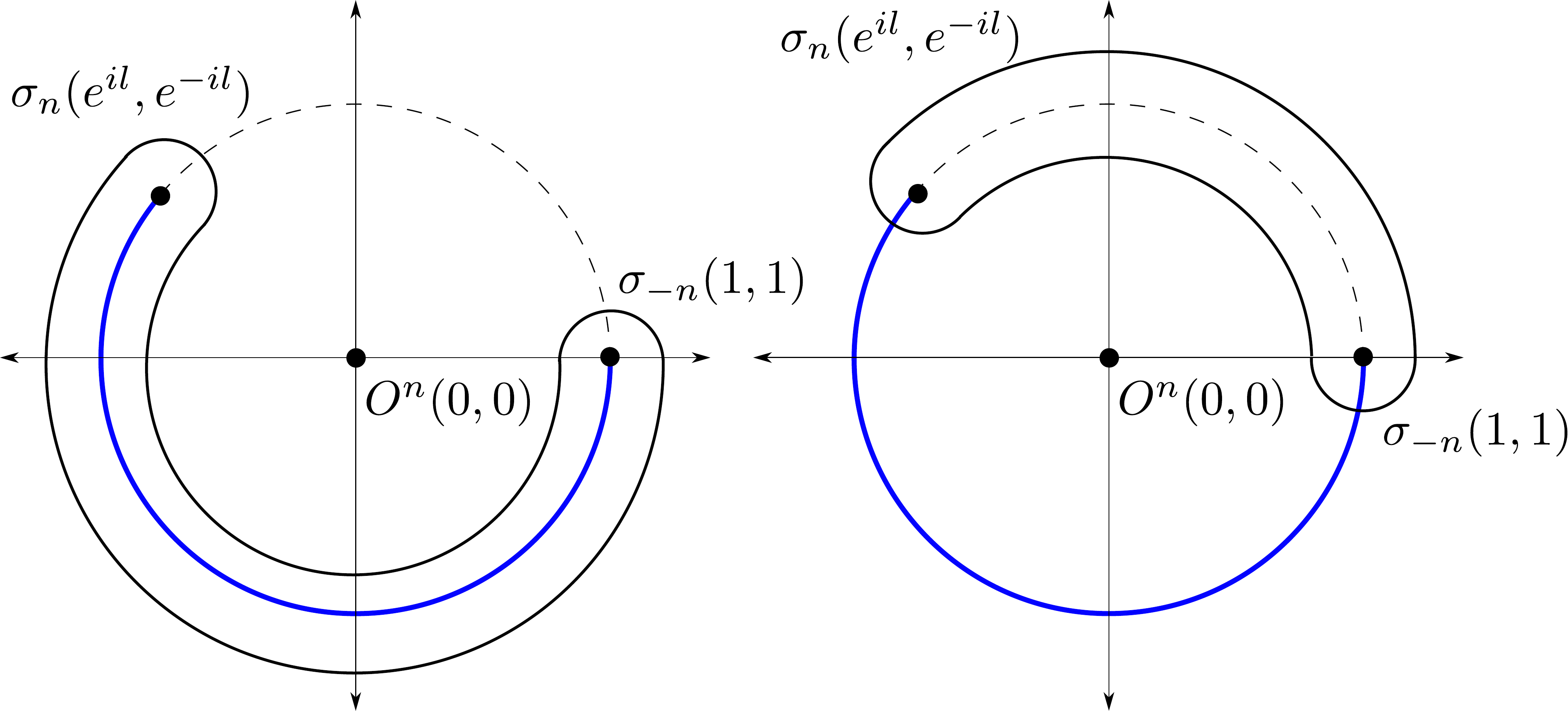}
\caption{Two different possible channels for computing the OPE between the two twist operators. The identity block contribution depends sensitively on the chosen channel. The identity block in the channel taken in the right diagram is more dominant than that of the left, and well approximates the four-point function. Dominance switches across $l = \pi$.}\label{channel}
\end{center}
\end{figure}

The resolution of this issue was discussed by \citep{Asplund:2014coa} where they point out that the identity block contribution in \ref{renyin} is not analytic. In particular it is not invariant under $l \rightarrow l + 2 \pi$. Due to this monodromy, the result is sensitive to how $\sigma_{n}(z,\bar{z})$ is wound around the origin where $\mO^n$ is located. Since there is more than one way to get to any point on the unit circle, there can be many different identity block `channels'. \citep{Asplund:2014coa} notes, however, that since the exact four point function is analytic, the dominant identity block channel must be equivalent to any subdominant identity block channel plus contributions from other non-identity blocks. Thus, the four point function is well approximated by the dominant identity block contribution across all channels. In this case, this is the channel which involves no winding around the origin and is taken along an arc of angle less than $\pi$. This is shown in figure \ref{channel}. With this understanding the Renyi entropy and, thus, the von-Neumann entropy are both symmetric under $l \leftrightarrow 2 \pi - l$.

\begin{figure}
\begin{center}
\includegraphics[height=6cm]{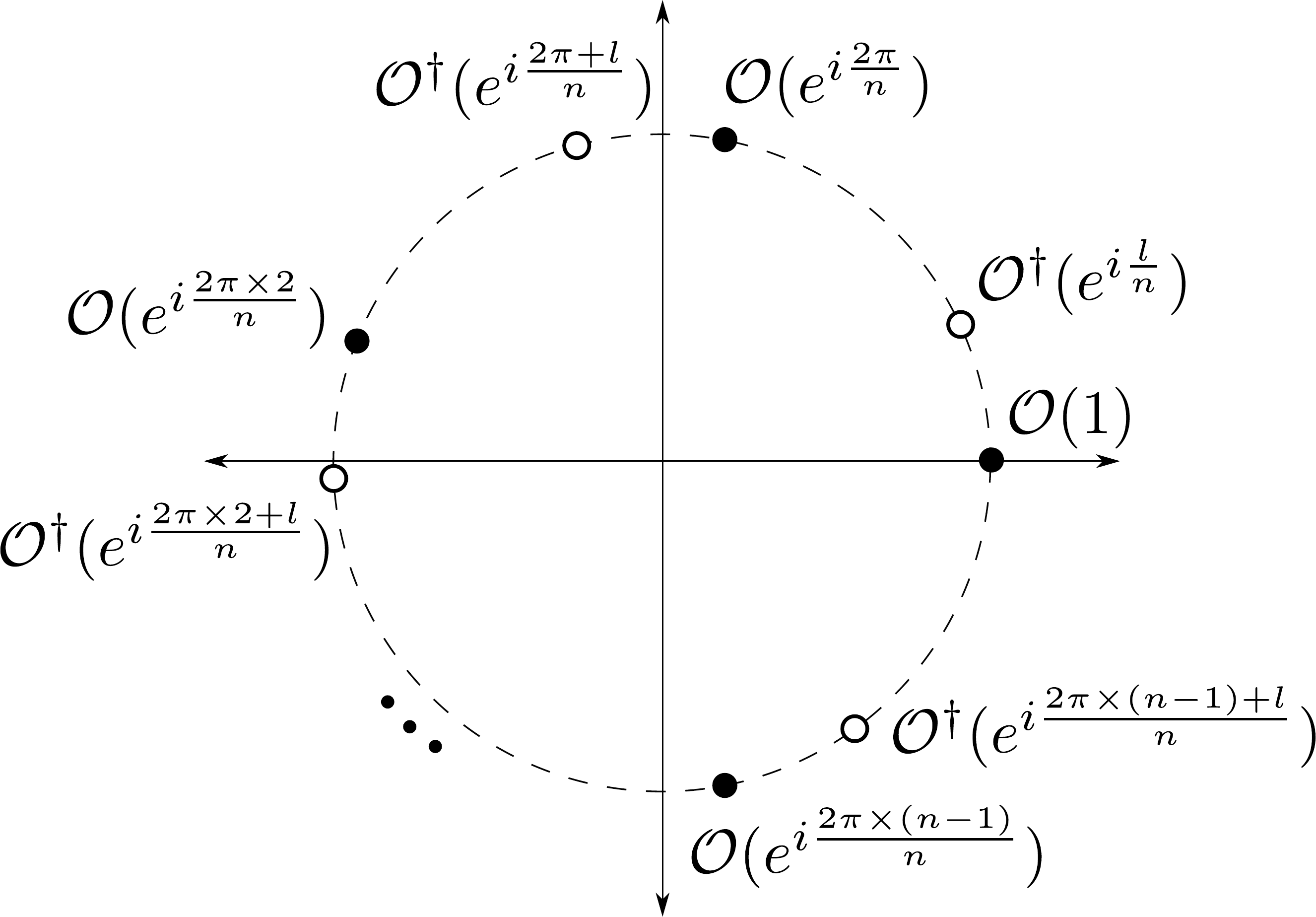}
\caption{The replicated density matrix is represented after uniformization as a 2$n$ point function of $\mO$ and $\mO^{\dagger}$ on the unit circle. The dark circles represent $\mO$ insertions while the hollow circles represent $\mO^\dagger$ insertions.}\label{opcircle}
\end{center}
\end{figure}

\begin{figure}
\begin{center}
\includegraphics[height=6cm]{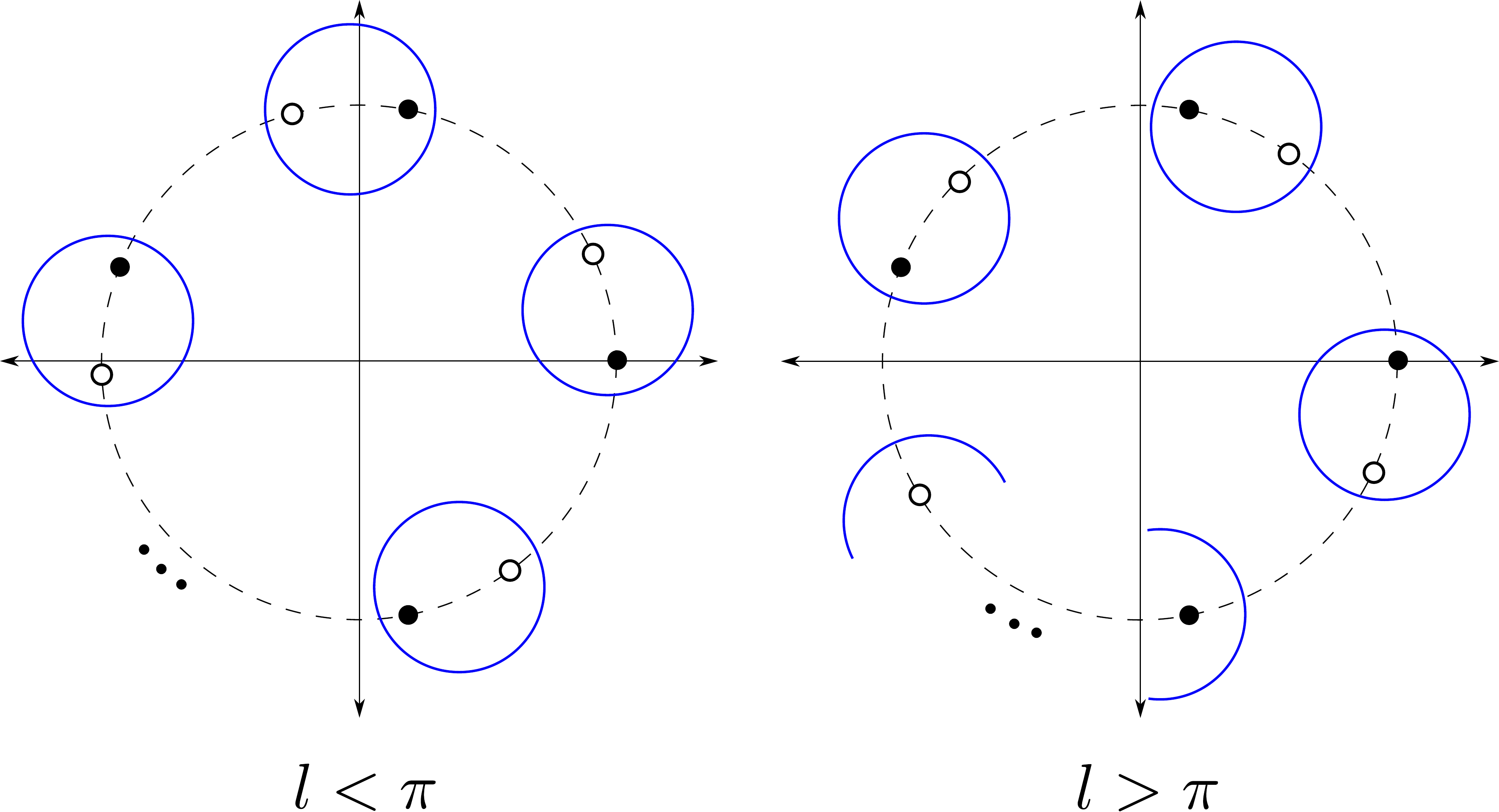}
\caption{The dominant identity block OPE channel for computing the replicated density matrix. The blue circles indicates how the OPE expansion is taken. The locations of $\mO^{\dagger}$ (the hollow circles) move as $l$ is changed. We see that the operator pairing switches at $l = \pi$.}\label{uchannel}
\end{center}
\end{figure}

There is actually a clearer way to see that \ref{4pnt} is manifestly symmetric under $l \leftrightarrow 2 \pi - l$. Consider performing a uniformizing coordinate transformation,
\begin{align}
w = \left[ \left(   {z - 1 \over z - e^{i l}}   \right) e^{i l}   \right]^{1/n}, \label{map}
\end{align}
that removes the twist operators and puts all the operators on a single complex plane. Under this transformation the coordinates map to
\begin{align}
&z \rightarrow 0 \ \ : \ \ w \rightarrow e^{i 2 \pi k\over n} \\
&z \rightarrow \infty \ \ : \ \ w \rightarrow e^{{i 2 \pi k\over n }  + {i l \over n}}
\end{align}
for $k$ an integer $\in [0,n)$. $k$ labels which branch an operator came from. In this coordinate system the four-point function becomes, up to a proportionality constant that depends on $l$ and is symmetric under $l \leftrightarrow 2\pi - l$, the following
\begin{align}
\langle 0 |\mO(1) \mO^{\dagger}(e^{i{ l \over n}}) \mO(e^{i { 2 \pi \over n} }) \mO^{\dagger}(e^{ i { 2 \pi + l\over n} }) ... \mO(e^{i { 2 \pi (n - 1) \over n} }) \mO^{\dagger}(e^{i { 2 \pi (n - 1)  + l \over n}})     |  0  \rangle.
\end{align}
This is a $2n$ point function of $\mO$'s located at $e^{i {2 \pi k \over n}}$ and $\mO^{\dagger}$'s placed in between at $e^{i {2 \pi k + l \over n}}$. This is shown in figure \ref{opcircle}. This representation makes it clear that the result will be symmetric under $l \leftrightarrow 2 \pi - l$. When $l < \pi$ the dominant contribution will be from the identity block taken in the channel $\mO (e^{i {2 \pi k \over n}}) \rightarrow \mO^{\dagger} (e^{i { 2 \pi k  + l \over n}})$. And When the case of $l > \pi$, the dominant contribution comes from the $\mO(e^{i {2 \pi (k+ 1) \over n}}) \rightarrow \mO^{\dagger} (e^{i { 2 \pi k  + l \over n}})$ channel. These two different channels are represented in figure \ref{uchannel}.

To conclude, the entanglement entropy of an interval in a heavy state of zero spin is given by
\begin{align}
S &= {c \over 3} \ln \left[ {\beta \over \pi \epsilon} \sinh\left( {l \pi \over \beta} \right)   \right], & l < \pi\\
&= {c \over 3} \ln \left[ {\beta \over \pi \epsilon} \sinh\left( {(2 \pi - l) \pi \over \beta} \right)   \right],  & l> \pi.
\end{align}
As noted in \cite{Asplund:2014coa}, this result can be obtained from the bulk using the RT prescription but without imposing the homology constraint. It is not actually clear what imposing this constraint would mean given that the interior of a black hole microstate is not really well understood.

And finally, one can also extract the answer for a light state, $h/c \rightarrow 0$ as $c \rightarrow \infty$, by simply continuing $h/c \rightarrow 0$. In this limit $\beta \rightarrow 2\pi i$ giving
\begin{align}
S = {c \over 3} \ln \left[ {2 \over  \epsilon} \sin \left( {l  \over 2} \right)   \right]
\end{align}
which works for all $l$.

\section{Entanglement Entropy for Superpositions of Semi-Classical States}

We present in this section the computation of entanglement entropy for states dual to macroscopic superpositions of semi-classical geometries. We focus mainly on two classes of such states: superpositions of pure one-sided states considered in section \ref{example} and superpositions of thermofield doubles of different temperatures. This will mostly be a summary and the explicit details will be left to appendix \ref{AppSuper}.

\subsection{Superpositions of One-Sided AdS Spacetimes}
\label{onesidedsuper}
Let us begin by considering superpositions of pure one-sided states constructed from the orthogonal basis
\begin{align}
\Big\{ \mO_i |0 \ran \Big\},
\end{align}
where $\mO_i$ are primary operators. States with low dimension correspond to perturbations of pure AdS, while those of high dimension correspond to black hole microstates.

We want to compute the entanglement entropy of an interval in states of the form
\begin{align}
| \Psi \rangle = \sum_{i = 1}^M \alpha_i \mO_i(0,0) | 0 \rangle \equiv \Psi |0\rangle, \label{pure}
\end{align}
where $\mO_i$ are orthogonal primary operators. Following the techniques of section \ref{review}, we can compute the entanglement entropy of an interval using the replica trick. Just as before, we need to compute the replicated density matrix of the interval. This is given by
\begin{align}
tr \rho^n &= \langle 0 | (\Psi^\dagger(\infty))^n \sigma_n(1,1) \sigma_n(z,\bar{z})   (\Psi(0,0))^n      | 0 \rangle \\
&=  \langle 0 |\left(\sum_i \alpha_i^* \left(\mO_i(0)\right)^\dagger\right)^n \sigma_n(z,\bar{z}) \sigma_{-n}(1,1) \left(\sum_i \alpha_i \mO_i(0)\right)^n     | 0  \rangle \\
&= \sum_{\substack{a_1, ..., a_{M} = 0 \\ b_1, ..., b_{M} = 0}}^n    \alpha_1^{ a_1} ... \alpha_{M}^{ a_{M}} \alpha_1^{* b_1} ... \alpha_{M}^{ * b_{M}}   \langle 0 | (\mO_1^\dagger)^{b_1} ... (\mO_{M}^\dagger)^{b_{M}}  \sigma_n(z,\bar{z}) \sigma_{-n}(1,1) \mO_1^{a_1} ... \mO_{M}^{a_{M}}  | 0 \rangle \label{tracerhonpure}
\end{align}
where
\begin{align}
\mO_1^{a_1} ... \mO_{M}^{a_{M}} \equiv \overbrace{\mO_1  \otimes...\otimes \mO_1}^{a_1} \otimes \dots \otimes \overbrace{\mO_M  \otimes...\otimes \mO_M}^{a_M} + \ \left( {n! \over a_1! ... a_M!} - 1 \right) \ \mathrm{permutations} \label{perm}
\end{align}
with the condition that $\sum_{i = 1}^M a_i = \sum_{i = 1}^M b_i  = n$. These are orbifold symmetric primary operators belonging to the orbifold CFT on $\mathbb{C}^n$.

The replicated density matrix \ref{tracerhonpure} is thus a sum of four point functions of heavy operators for $n>1$. We are interested in computing this quantity for a holographic theory, so we assume the all the four point functions are well approximated by their identity block contribution. Restricting to the identity block in the $t$-channel offers an immediate simplification of the above expression. Since the identity block can only appear in the expansion of two non-orthogonal operators of the same dimension, only terms with $a_i = b_i$ contribute. Thus, \ref{tracerhonpure} reduces to
\begin{align}
tr \rho^n &= \sum_{a_1, ..., a_{M} = 0}^n    |\alpha_1|^{2  a_1} ... |\alpha_{M}|^{2 a_{M}}   \langle 0 | \mO_1^{\dagger {a_1}} ... \mO_{M}^{ \dagger {a_{M}} }  \sigma_n(z, \bar{z}) \sigma_{-n}(1,1) \mO_1^{a_1} ... \mO_{M}^{a_{M}}  | 0 \rangle. \label{idenden}
\end{align}
This expression is also symmetric under $l \leftrightarrow 2 \pi - l$ for the very same reasons \ref{replden} is as explained in figure \ref{uchannel}. Let us see how this works explicitly. Let us call the terms where any $a_i = n$ the `diagonal' terms and everything else the `off-diagonal' terms.

It is clear that the diagonal terms have the exact form as \ref{replden}, and so this symmetry follows by the same reasoning. There is an interesting twist for the off-diagonal terms. The different operator orderings in \ref{perm} for the operator and its complex conjugate pair up in just the right way as $l$ is changed. For simplicity, let us focus on the $n = 2$ case for a superposition of only two primary states. The off-diagonal term of the $n = 2$ replicated density matrix is given by
\begin{align}
tr \rho^2_{OD} &\propto \langle 0 | \big(\mO_1^\dagger \otimes \mO_{2}^\dagger + \mO_2^\dagger \otimes \mO_{1}^\dagger \big) \sigma_2(z, \bar{z}) \sigma_{-2}(1,1) \big( \mO_1\otimes  \mO_{2} + \mO_2\otimes  \mO_{1} \big)  | 0 \rangle \\
&\propto \langle 0 | \mO_1^\dagger \otimes \mO_{2}^\dagger  \sigma_2\sigma_{-2}  \mO_1\otimes  \mO_{2}   | 0 \rangle  + \langle 0 | \mO_2^\dagger \otimes \mO_{1}^\dagger  \sigma_2 \sigma_{-2}  \mO_2\otimes  \mO_{1}   | 0 \rangle \nonumber  \\
& \ + \langle 0 | \mO_1^\dagger \otimes \mO_{2}^\dagger  \sigma_2\sigma_{-2}  \mO_2\otimes  \mO_{1}   | 0 \rangle  + \langle 0 | \mO_2^\dagger \otimes \mO_{1}^\dagger  \sigma_2 \sigma_{-2}  \mO_1\otimes  \mO_{2}   | 0 \rangle.
\end{align}
Notice the difference in the operator orderings in the last equation. After uniformizing, we find that the channel which expands $\mO_1^\dagger$ with $\mO_1$  and  $\mO_2^\dagger$ with $\mO_2$  in the first two terms expands $\mO_1^\dagger$ with $\mO_2$  and  $\mO_2^\dagger$ with $\mO_1$ in the latter two, and vice versa. Thus, the identity block exists in either the first pair of terms or the second and not together. We are forced to apply the same channel for all the terms since that choice is inherited from picking a channel of the orbifold symmetric operators in the four point function before breaking it up into its components. It turns out that the identity block from the first pair of terms dominates for $l < \pi$ and from the second pair for $l > \pi$. This exchange ensures the result has the required symmetry. The same line of reasoning applies for arbitrary $n$ and superposition.

To finally compute the entropy, we need to evaluate the four point functions appearing in \ref{idenden}, and then preform the sum over the $a_i$'s. As discussed previously, these four point functions are not known in closed form for $n > 1$, except as a perturbative expansion in $l$. We evaluate this expression with the following series of manipulations:
\begin{enumerate}
\item Consider first the different quantity\begin{align}
tr \rho^n_m \equiv \sum_{a_1, ..., a_{M} = 0}^m    |\alpha_1|^{2  a_1} ... |\alpha_{M}|^{2 a_{M}}   \langle 0 | \mO_1^{\dagger {a_1}} ... \mO_{M}^{ \dagger {a_{M}} }  \sigma_n(z, \bar{z}) \sigma_{-n}(1,1) \mO_1^{a_1} ... \mO_{M}^{a_{M}}  | 0 \rangle.
\end{align}
This differs from \ref{idenden} in the upper limit of the $a_i$ sums. It is clear that $\lim_{m \rightarrow n} tr \rho^n_m = \tr \rho^n$.

\item Take the limit of $n$ approaching 1, holding $m$ fixed,  where we know the explicit forms of the four point functions appearing in the sum.

\item It turns out that even after plugging in these forms, it is still not easy to perform the sum over $a$. We get around this by first performing an expansion in $l$ and then do the $a$ sum term by term.

\item Then take the limit as $m \rightarrow n$ and act with $\lim_{n \rightarrow 1} \partial_n$ to obtain the entropy.

\item Finally, resum the series in $l$.
\end{enumerate}
We believe this procedure gives the correct entanglement entropy based on the following two strong pieces of evidence. One, it reproduces the result from the perturbative expansion of the identity block in the size of the interval. Two, it maintains the requirement that $\lim_{n\rightarrow1}\tr \rho^n = 1$. The details of the calculation are presented in appendix \ref{AppSuper}.

The result we find is that the identity block contribution to the entanglement entropy in the superposition \ref{pure} is exactly
\begin{align}
S = \sum_{i = 1}^M |\alpha_i|^2 S_i
\end{align}
where $S_i$ is the entanglement entropy of an interval in the state $\mO_i |0\ran$. This will be a good approximation to the entropy as long as the identity block contribution to the replicated density matrix remains dominant. However, as $M$ is increased the number of non-identity block contributions proliferates faster than the identity block terms; there are $M^{2 n}$ terms of the former and $M^n$ of the latter. The magnitude of the individual terms from the identity block is larger than a typical non-identity block term by a factor of $e^{\# n c}$. Thus, we expect that the identity block approximation fails once $M \sim e^{O(c)}$.

\subsection{Superpositions of Eternal Black Holes}
\label{subsec:supereternal}
Next, let us consider superpositions of thermofield double states of different temperature. These are states defined on a product Hilbert space of two CFTs each living on $S \times R$, and are dual to macroscopic superpositions of eternal black holes of different masses.  Such states are given by
\begin{align}
| \Psi \rangle = \sum_{i = 1}^M \alpha_i | \beta_i \rangle,
\end{align}
where
\begin{align}
| \beta_i \rangle = {1 \over \sqrt{Z(\beta_i)}} \sum_E e^{- \beta_i E/2} | E \rangle_L | E \rangle_R
\end{align}
and $Z(\beta) = e^{\pi^2 c \over 3 \beta}$ is the partition function of the theory. This state corresponds to a bulk superposition of eternal black holes of different mass $M_i = \pi^2 c/3 \beta_i^2$.

Say we want to compute the entanglement entropy of the right CFT. For a single TFD this computes the Bekenstein-Hawking entropy of the dual black hole. To obtain the entropy in the superposition, we first compute the reduced density matrix of the right CFT and find
\begin{align}
\rho_R = Tr_L | \Psi \rangle \langle \Psi | = \sum_{i = 1}^M |\alpha|^2 \rho_i + \sum_{i \neq j = 1}^M 2 Re \left( \alpha_i^* \alpha_j \right) \sqrt{\rho_i \rho_j},
\end{align}
where $\rho_i = e^{- \beta_i E}/Z(\beta_i)$. Immediately, the entanglement entropy is computed by
\begin{align}
S_R &= - Tr \rho_R \ln \rho_R \\
&= - \int dE D(E) \rho_R \ln \rho_R \\
& = - \int dE D(E) \left(\sum_{i = 1}^M |\alpha_i|^2 \rho_i + \sum_{i \neq j = 1}^M 2 Re \left( \alpha_i^* \alpha_j \right) \sqrt{\rho_i \rho_j} \right) \ln \rho_R \label{tfdentropy}
\end{align}
where $D(E)$ is the density of states, which for a holographic CFT on a cylinder scales as $e^{2 \pi \sqrt{c E\over 3}}$ for large $E$. This expression can be evaluated term by term via saddle point. Focusing on a term in the first sum of the above expression, we find that the saddle point evaluates to
\begin{align}
\int dE D(E) \rho_i \ln \rho_R \sim \ln \rho_R|_{E\rightarrow \pi^2 c/ 3 \beta_i^2}.
\end{align}
It can be easily checked that $\rho_R \rightarrow |\alpha_i|^2\rho_i$ as  $E\rightarrow \pi^2 c/ 3 \beta_i^2$;  $\rho_i$ is always picked out as the dominant term in the logarithm. Thus, the contribution of these terms to the entropy is given by
\begin{align}
- \sum_{i = 1}^M |\alpha_i|^2\int dE D(E) \rho_i \ln \rho_R &\sim - \sum_{i = 1}^M |\alpha_i|^2 \rho_i \ln |\alpha_i|^2 \rho_i \\
&\sim  \sum_{i = 1}^M |\alpha_i|^2 S(\rho_i) - \sum_{i = 1}^M |\alpha_i|^2 \ln |\alpha_i|^2.
\end{align}
The first term is simply the average of the entropies of the different branches of the wavefunction, while the second is a classical Shannon entropy known also as the entropy of mixing \cite{2006PhRvL..97j0502L}.

The second sum in \ref{tfdentropy} can also be evaluated via saddle point, and we find
\begin{align}
\int dE D(E)\sqrt{\rho_i \rho_j} \ln \rho_R \sim e^{- {4 \pi^2 c\over 3} \left( {1\over 8 \beta_i} + {1\over 8 \beta_j}  - {1 \over 2 (\beta_i + \beta_j)}\right)}\ln \rho_R|_{E\rightarrow 2 \pi^2 c/3 ( \beta_i + \beta_j)^2} \label{offTFD}
\end{align}
which is exponentially suppressed in $c$ unless $\beta_i = \beta_j$. This result essentially follows from the near orthogonality of the thermofield double states of different temperature; their overlap is suppressed by the same exponential factor.

Putting these results together, we find that the entanglement entropy of the right CFT is
\begin{align}
S_R = \sum_{i=1}^M |\alpha_i|^2 S_i - \sum_{i = 1}^M |\alpha_i|^2 \ln |\alpha_i|^2. \label{tfdresult}
\end{align}
Thus, the entropy averages up to the entropy of mixing term. The entropy of mixing can be at most $\ln M$, so as long as $M$ is much less than $e^{O(c)}$, the entropy of mixing term may be neglected and the entropy averages. Once $M$ is of order $e^{O(c)}$, the entropy of mixing term can in principle compete with the average term; additionally, many of the approximations made in reaching \ref{tfdresult} become unreliable when $M \sim e^{O(c)}$. So we do not expect and have no evidence that the entropy averages in this regime.

\section{Linearity vs Homology}
\label{linhom}

We showed in the previous section that, to leading order in $c$, the entanglement entropy of an interval in states dual to macroscopic superpositions of a small number of distinct classical geometries is given by the average of the entropy in each branch of the wavefunction, thus confirming the prediction \ref{pred}. This is consistent with the statement that the entropy is approximately represented as the expectation value of a linear operator. This linear operator must have small off-diagonal matrix elements between semi-classical states, consistent with the structure of the area operator. As before, all statements are valid for superpositions of much fewer than $e^{O(c)}$ semi-classical states.

Moreover, we identified a new correction to the RT formula, the entropy of mixing, which we expect to appear when the density matrices of the CFT subregion, and its complement, in the different branches are distinguishable. In the regime where the leading contribution to RT is the average of the areas of the different branches of the wavefunction, this mixing term is subleading as compared to the area term.

It seems thus far that the leading contribution of the RT proposal is given by the expectation value of a linear operator, namely the area operator. However, in this section we identify another nonlinearity associated with the area contribution which arises when considering $e^{O(c)}$ states but which manifests in different way. In contrast to the failure of nonlinearity discussed in the previous section, this contribution we will be able to compute exactly.

\subsection{A Failure of Linearity: Homology}

In order to see this nonlinearity, we restrict the RT formula to the area term which is always the leading order in $c$ contribution in any semi-classical state. For simplicity we continue to work in the context of 1+1 holographic CFTs. Then, the prescription for computing the entanglement entropy of  an interval ${\cal I}$ in the state $|\Psi \rangle$ is
\begin{align}
S\big( {\cal I},  | \Psi \rangle \big) = \langle \Psi | \hat{\cal A}_{\cal I} | \Psi \rangle.
\end{align}

We saw in the previous section that when dealing with single sided pure states the entanglement entropy truly behaved like the expectation value of $\hat{\A}_{\cal I}$ within subspaces of semi-classical states spanned by $\big\{ \mO_i |0 \ran \big\}$\footnote{Recall, we only showed this for primary states where the identity block dominates. We assume it continues to hold for descendant states. However, we cannot rule out the possibility of states for which the identity block does not dominate, but we expect these to be rare at high energies.} and of dimension much less than $e^{O(c)}$. These are pure states of one CFT on one connected manifold, specifically $S^1$. One can ask whether this same operator continues to work for mixed states of this CFT, or more specifically, for pure states of two copies of the same CFT. We will focus on the latter case of a CFT living on $S^1_L \cup S^1_R$, which we label as left, $L$, and right, $R$. The question now is whether $\hat{\A}_{\cal I}$ applied to, say, the right CFT correctly computes the entanglement entropy of an interval on states composed of the basis elements $\big\{  \mO^L_i   \ |0 \ran_L \otimes \mO^R_j \ |0 \ran_R \big\}$. If the leading contribution of RT is truly represented by a linear operator then this must be the case.

It is clear that it would do so for any single element of this basis, and also for any superposition that produces a pure density matrix for both CFTs. To see the failure of linearity, we need to consider a highly entangled state between the two CFTs. The most convenient such state to consider is the thermofield double which contains order $c$ entanglement between the two CFTs. We choose one where the inverse temperature $\beta$ is small enough such that the dominant configuration is an eternal black hole. Using the operator $\hat{\A}_{\cal I}$, the entropy of an interval on the right CFT is
\begin{align}
S({\cal I}, | \beta  \rangle ) &= \langle \beta | \hat{\A}_{\cal I} | \beta \rangle \\
 &= \sum_E {e^{- \beta E} \over Z(\beta)} \langle E | \hat{\A}_{\cal I}| E \rangle. \label{shattfd}
\end{align}
The first equality is simply the application of the RT formula. The second comes from the fact that $\hat{\A}_{\cal I}$ is an operator purely on the right CFT as is suggested from the entanglement wedge reconstruction proposal discussed above in section \ref{entwedgerecon}. Equation \ref{shattfd} says that the entropy in the thermal state is simply the thermal average of the entropy in the eigenstates. We can evaluate this sum via saddle point methods while keeping in mind that the area operator is a coarse operator and will not shift the saddle point to leading order in $c$ as discussed in section \ref{subsec:areaop_predict}. We find that
\begin{align}
S({\cal I}, | \beta  \rangle ) &\approx \langle E_s | \hat{\A}_{\cal I}| E_s \rangle, \\
&\approx S({\cal I}, | E_s  \rangle ) \label{ssaddle}
\end{align}
where $E_s = \pi^2 c / 3 \beta^2$ is the average energy of the canonical ensemble at temperature $\beta$. Thus, we have found that the entanglement entropy in the thermal state can be approximated by that of the pure state at the average energy of that ensemble. The state $| E_s\rangle$ is a pure black hole of the right CFT whose exterior geometry agrees with that of the thermal state to leading order in $c$.

This result is immediately problematic; consider the situation where we are computing the entropy of the entire CFT, or ${\cal I} = 2\pi$. This implies
\begin{align}
S(2 \pi , | \beta  \rangle ) \approx S(2 \pi, | E_s  \rangle ) = 0
\end{align}
which is obviously wrong! This should compute the entanglement between the two CFTs in the thermal state which is proportional to $c$, reproducing the holographic result of computing the area of the eternal black hole. This issue is very reminiscent of the earlier objection  using qubits discussed in the introduction; the entanglement entropy operator which computes the entropy of the entire CFT is the zero operator when constructed in a basis of pure states. Note also that this is different from the problem of cross terms in the area operator adding up and changing the answer when there are too many states in the superposition. The reason for this distinction is that the thermal density matrix is diagonal and thus the cross terms  $\lan E' | \hat{\mathcal{A}} | E \ran$ do not appear. We will discuss this issue and its relation to the CFT calculation more carefully in the next section.

Surprisingly, however, the formula does not fail for all interval sizes. Let us consider the bulk prescriptions for computing the entropy as a function of the size of the interval for the thermal state and the pure state. Starting with a small interval, we find that formula \ref{ssaddle} gives the correct answer to leading order in $c$ up until ${\cal I} = \pi$. The discrepancy begins as soon as $\mathcal{{\cal I} > \pi}$ and gets worse as we make the interval larger. As noted, while \ref{ssaddle} falls down to zero, as it must, the thermal answer saturates at the thermal entropy $ 2 \pi^2 c /3 \beta$.

\begin{figure}
\begin{center}
\includegraphics[height=5cm]{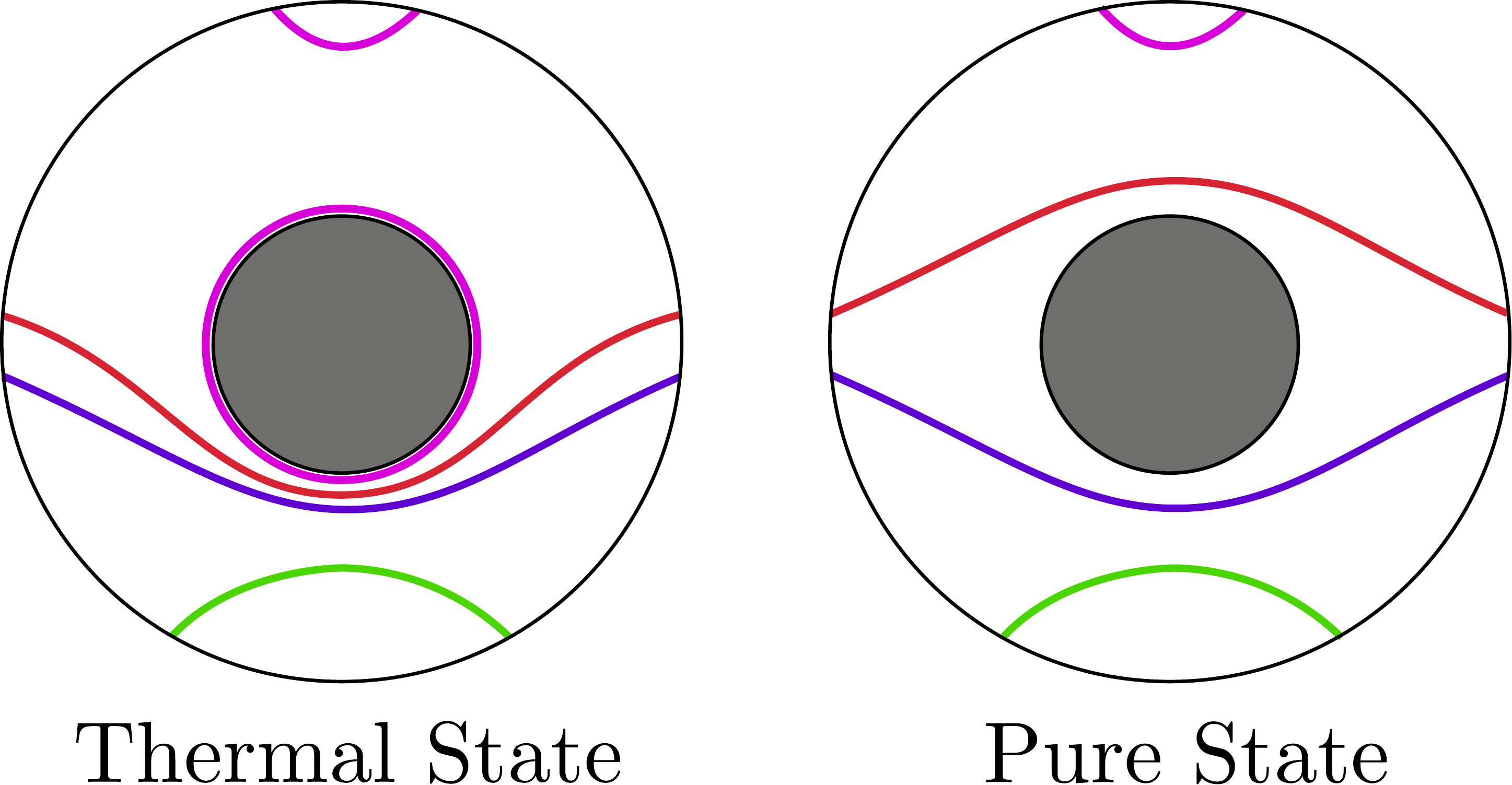}
\caption{Minimal area surfaces which compute the entanglement entropy of various intervals of the boundary CFT. Entropy of intervals smaller than $\pi$ are the same for both a pure and eternal block hole, and are given by green and blue curves. The two cases begin to differ once the interval is larger than $\pi$, as those are given by different bulk surfaces as shown by the red and magenta curves. We note that the difference for intervals that cover almost the entire boundary is exactly the black hole entropy.}\label{stoned}
\end{center}
\end{figure}

The holographic reason for this discrepancy is clear, and is presented pictorially in figure \ref{stoned}. From the bulk perspective, the difference stems from the differing bulk prescriptions for picking out the minimal area extremal surface in the single sided black hole geometry versus the two sided eternal black hole. Recall that these geometries agree in the exterior of the black hole. As shown in figure \ref{stoned}, the extremal surface that computes the entropy for small intervals is the same for both cases up until ${\cal I} = \pi$. Beyond this point, the extremal surface for the single sided case jumps across to the other side of the black hole, while it remains on the same side for the thermal case. Even though the surface on the other side has smaller area than the surface on the same side, the homology constraint forces the surface of thermal case to stay on the same side. As previously discussed in section \ref{example} and \cite{Asplund:2014coa}, it is not clear what it means to impose the homology constraint in the pure case as there might not be a geometric interior to these black holes \cite{Almheiri:2012rt,Almheiri:2013hfa}. Nevertheless, the CFT result requires the jump to the other side. This can be interpreted loosely as not imposing the homology constraint in the pure case.

We conclude that there is not a single entropy operator $\hat{\A}_{\cal I}$ which gives the correct entropy for pure and highly mixed states for all intervals ${\cal I}$. From a bulk perspective, the nonlinearity was introduced by the requirement of imposing the homology constraint in one case but not in the other. One can thus think of the homology prescription as specifying the set of surfaces $\hat{X}$ that we are allowed to extremize the area operator over. Thus, the homology constraint precludes the entropy from being the expectation value of a single linear operator defined on only one CFT.

We note that homology being the source of nonlinearity of entanglement entropy is very reminiscent of the recently discussed `wormhole' operator that measures whether two separate AdS bulk spacetimes are connected via an Einstein-Rosen bridge. This is clearly a nonlinear property of a state since the thermofield double, while dual to a wormhole, is a superposition\footnote{This connection between superpositions and topology change has also been recently investigated in \cite{Berenstein:2016pcx} in the context of LLM geometries.} of states with manifestly no geometric connection \cite{Papadodimas:2015xma}.\footnote{This conclusion has been argued against in \cite{Marolf:2012xe} which posits the necessity of `superselection information' that determines whether the TFD is dual to a wormhole or a pair of disconnected black holes.}

\subsection{The Source of the Homology Constraint in the CFT}

From the bulk perspective, the discrepancy found in the previous section was due to imposing the homology constraint. The considered thermal state is a two-sided superposition of $e^{O(c)}$ states; the number of states is actually infinite, but the relevant terms which dominate the canonical ensemble are those with energy roughly the average energy at the considered temperature and number around  $e^{O(c)}$. Therefore, we see that linearity fails once we have a large number of terms in the superposition. However, in contrast to the previous issue of non-identity block contributions becoming important, we will see that the homology constraint can be explained via exchange of identity block channel dominance.

Let us compare the entanglement entropy computation of two states with the same bulk dual, at least from one side. First, consider an approximate form of the thermofield double. The TFD state
\begin{align}
|TFD \rangle = {1 \over \sqrt{Z(\beta)}} \sum_E e^{- \beta E \over 2} | E \ran_L |E\ran_R
\end{align}
can be approximated by terms within an energy shell of  width $O(c^0)$ around the average energy, $E_s = \pi^2 c /3 \beta^2$. This defines a microcanonical ensemble.  Let us assume $\beta$ is small enough to be above the Hawking-Page transition, thus each term in this state is dual to a large black hole in AdS. We can then estimate the number of terms in the considered energy shell to be given by the Cardy formula, $e^{2 \pi \sqrt{c E_s \over 3}} = e^{2 \pi^2 c/3  \beta}$. This approximate state is
\begin{align}
|\widetilde{TFD} \rangle &= {1 \over Z(\beta)} \sum_{i = 1}^{e^{2 \pi \sqrt{c E_s / 3}}} e^{- \beta E_s \over 2} | E_i \rangle_L | E_i \rangle_R \\
&= e^{- {\pi^2 c \over 3 \beta}} \sum_{i = 1}^{e^{2 \pi^2 c/3 \beta}} \mO_i^L\otimes \mO_i^R | 0 \rangle_L | 0 \rangle_R \label{approxtildetfd}
\end{align}
where we take the $\mO_i^{L,R}$ to be primary operators of dimension roughly $E_s$. The restriction to primary operators is a further approximation, since the number of descendant states in the considered energy shell is an order one fraction of the total number of states. However, this approximate state is expected to be accurate when studying coarse-grained observables, namely those that satisfy ETH. As we will momentarily show, this state reproduces the RT result of the entanglement entropy of an interval in a state dual to an eternal black hole.

We will actually first consider a truncated version of \ref{approxtildetfd} to any $M$ terms,
\begin{align}
| \mathrm{Mixed} \ran= {1 \over \sqrt{M}} \sum_{i = 1}^{M} \mO_i^L\otimes \mO_i^R | 0 \rangle_L | 0 \rangle_R. \label{hommixed}
\end{align}
The specific choice will not matter since all of these operators have roughly the same dimension. The state we want to compare this to is a pure state on the right CFT constructed from the right operators appearing in \ref{hommixed}. This is
\begin{align}
| \mathrm{Pure} \ran = {1 \over \sqrt{M}} \sum_{i = 1}^{M}  \mO_i^R  | 0 \rangle_R. \label{hompure}
\end{align}
The bulk dual of this state is a pure black hole, or a microstate, whose exterior geometry is given by that of BTZ. Recall in our discussion below equation \ref{psiareapsi}, such a state is not atypical enough from the perspective of coarse observables. Both states \ref{hommixed} and \ref{hompure} describe the same right exterior geometry.

Let us compute the entanglement entropy of an interval in these states. Their replicated density matrices are
\begin{align}
tr \rho_{\mathrm{Pure}}^n  &=M^{- {n }} \! \sum_{a_1, ..., a_{M} = 0 }^n   { }_R\langle 0 | \mO_1^{\dagger{a_1}} ... \mO_{M}^{\dagger{a_{M}}}  \sigma_n(e^{i l},e^{- i l}) \sigma_{-n}(1,1) \mO_1^{a_1} ... \mO_{M}^{a_{M}}  | 0 \rangle_R  \label{puredensity}\\
tr \rho_{\mathrm{Mixed}}^n   &= M^{- {n }} \sum_{\substack{a_1, ..., a_{M} = 0 \\ b_1, ..., b_{M} = 0}}^n   { }_R\langle 0 |  \mO_1^{\dagger{b_1}} ... \mO_{M}^{\dagger{b_{M}}}   \sigma_n(e^{i l},e^{- i l}) \sigma_{-n}(1,1) \mO_1^{a_1} ... \mO_{M}^{a_{M}}  | 0 \rangle_R\nonumber \\
& \ \ \ \ \ \ \ \ \ \ \ \ \ \ \ \ \ \ \ \ \ \ \  \ \ \ \ \  \ \ \ \ \ \ \ \ \  \ \ \ \ \times {}_L\lan 0 |   \mO_1^{\dagger{b_1}} ... \mO_{M}^{\dagger{b_{M}}}   \mO_1^{a_1} ... \mO_{M}^{a_{M}}    |0 \ran_L \nonumber \\
 &\bcontraction[2ex]{=M^{- {n }} \! \sum_{a_1, ..., a_{M} = 0 }^n   { }_R\langle 0 |}{\mO_1^{\dagger{a_1}} ... \mO_{M}^{\dagger{a_{M}}} }{\sigma_n(e^{i l},e^{- i l}) \sigma_{-n}(1,1)}{\mO_1^{a_1} ... \mO_{M}^{a_{M}} }
=M^{- {n }} \! \sum_{a_1, ..., a_{M} = 0 }^n   { }_R\langle 0 | \mO_1^{\dagger{a_1}} ... \mO_{M}^{\dagger{a_{M}}}  \sigma_n(e^{i l},e^{- i l}) \sigma_{-n}(1,1) \mO_1^{a_1} ... \mO_{M}^{a_{M}}  | 0 \rangle_R,\label{mixeddensity}
\end{align}
where
\begin{align}
\mO_1^{a_1} ... \mO_{M}^{a_{M}} \equiv \overbrace{\mO_1  \otimes...\otimes \mO_1}^{a_1} \otimes \dots \otimes \overbrace{\mO_M  \otimes...\otimes \mO_M}^{a_M} + \ \left( {n! \over a_1! ... a_M!} - 1 \right) \ \mathrm{permutations},
\end{align}
and the contraction symbol between the operators indicates pairing of the same permutation. This is the only difference between the two replicated density matrices. Also, the pure replicated density matrix obtains the presented form only after restricting to terms with identity block contributions.

Let us compare the contributions to the entropy term by term, starting with the $a_i = n$ terms. These are equal in both cases and produce the result
\begin{align}
&l < \pi \ \mathrm{Channel}: \ \ \ S_{a_i = n}^{\mathrm{Pure}} = S_{a_i = n}^{\mathrm{Mixed}} =  \ln M + {c \over 3} \ln\left[{\beta \over \pi \epsilon} \sinh \left( {l \pi \over \beta}  \right) \right] \\
&l > \pi\ \mathrm{Channel}: \ \ \ S_{a_i = n}^{\mathrm{Pure}} = S_{a_i = n}^{\mathrm{Mixed}} =  \ln M + {c \over 3} \ln\left[{\beta \over \pi \epsilon} \sinh \left( {(2 \pi - l) \pi \over \beta}  \right) \right].
\end{align}
Just as in figure \ref{uchannel}, the $l < \pi$ channel corresponds to uniformizing and expanding the operators on $e^{i {2 \pi k \over n}}$ and $e^{i {(2 \pi + l)k \over n}}$ together, and the $l > \pi$ channel corresponds to expanding the operators on $e^{i {2 \pi k \over n}}$ and $e^{i {(2 \pi - l)k \over n}}$ together.

Things are a bit trickier for the $a_i \neq n$ terms. Let us do this channel by channel. Due to how the operators are arranged after uniformizing, the $l < \pi$ channel will only involve expansions of the form $\mO_i^\dagger \rightarrow \mO_i$ for both states, and will definitely have an identity block contribution. On the other hand, the states differ in their contribution in the $l > \pi$ channel. Since the permutations in the mixed case are matched, this channel will involve at least one expansion of the form $\mO_i^\dagger \rightarrow \mO_j$ with $i \neq j$, and will not receive an identity block contribution; orthogonal operators cannot fuse into the identity and its descendants. As for the pure case, the sum over permutations ensures there will always be a combination such that the $l > \pi$ channel expands $\mO_i^\dagger \rightarrow \mO_i$, and so will have an identity block contribution. Using the techniques of appendix \ref{AppSuper}, we find these contributions to be
\begin{align}
&l < \pi \ \mathrm{Channel}: \ \ \ S_{a_i \neq n}^{\mathrm{Pure}} = S_{a_i \neq n}^{\mathrm{Mixed}} =  - \ln M \\
&l > \pi\ \mathrm{Channel}: \ \ \ S_{a_i \neq n}^{\mathrm{Pure}}  = - \ln M, \   S_{a_i \neq n}^{\mathrm{Mixed}} = 0
\end{align}
Combining the contributions from both channels we get
\begin{align}
&S_{\mathrm{Pure}} = \mathrm{Min}\left(   {c \over 3} \ln\left[{\beta \over \pi \epsilon} \sinh \left( {l \pi \over \beta}  \right) \right]  \ ;  \ {c \over 3} \ln\left[{\beta \over \pi \epsilon} \sinh \left( {(2 \pi - l) \pi \over \beta}  \right)\right]   \  \right) \\
&S_{\mathrm{Mixed}} = \mathrm{Min}\left(   {c \over 3} \ln\left[{\beta \over \pi \epsilon} \sinh \left( {l \pi \over \beta}  \right) \right]  \ ;  \ \ln M + {c \over 3} \ln\left[{\beta \over \pi \epsilon} \sinh \left( {(2 \pi - l) \pi \over \beta}  \right)\right]   \  \right).
\end{align}
The minimizing prescription comes from the rule that the correct identity block approximation to the replicated density matrix is the one which dominates over all other identity block contributions across all channels.

\begin{figure}
\begin{center}
\includegraphics[height=5cm]{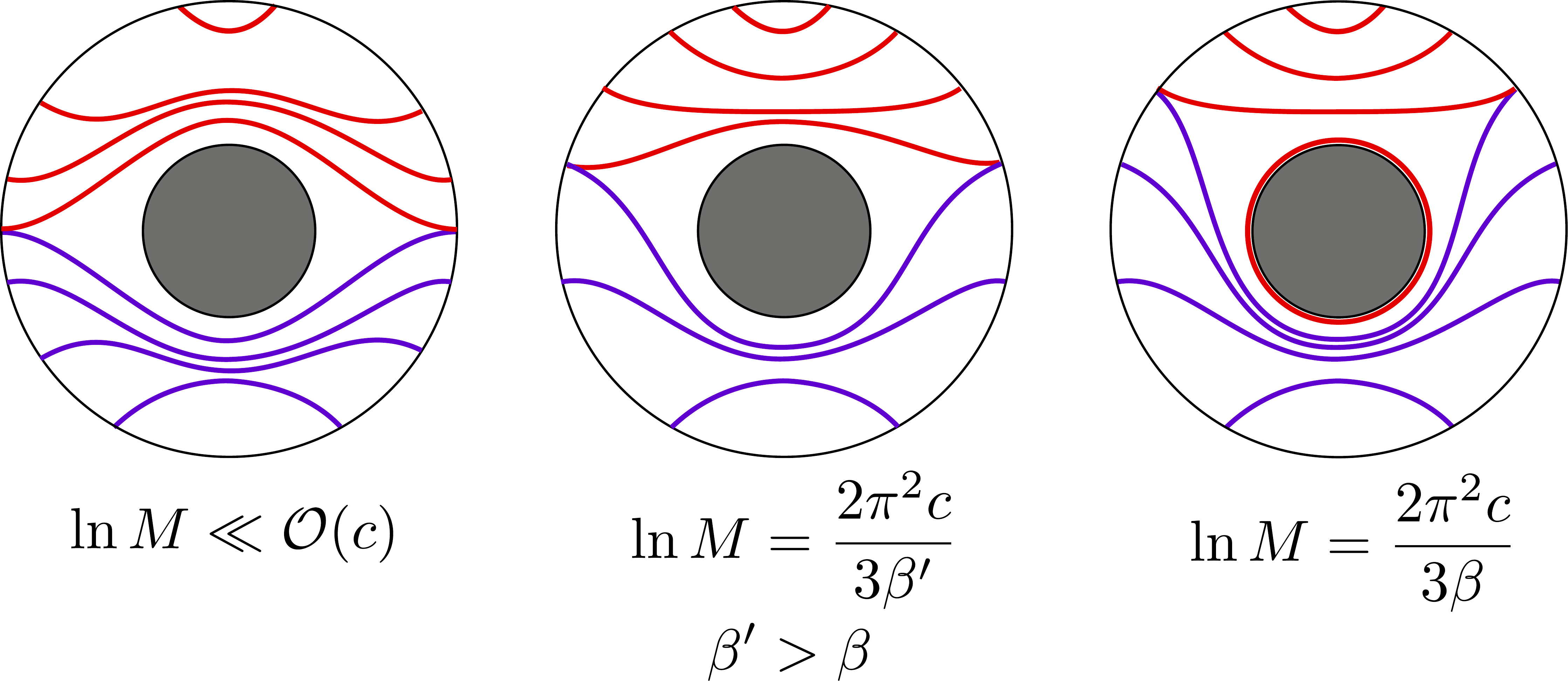}
\caption{The behavior of the RT surfaces as a function of $M$. When $\ln M \ll O(c)$ then we see that the transition occurs at $l = \pi$, just as in the pure state. For $\ln M = {2 \pi^2 c \over 3 \beta}$ attaining its maximum value, the RT surface transition occurs just as in the thermal state. The $\ln M$ piece computes the area the of the horizon. For the intermediate regime, where $\ln M = {2 \pi^2 c \over 3 \beta'}$ for $\beta' > \beta$, then the transition occurs somewhere in between. $\ln M$, in this case, does not compute the area of any surface in the exterior geometry.}\label{homologyM}
\end{center}
\end{figure}

Notice that both of these entropies have a discontinuous first derivative at some value of $l$. From the bulk perspective, this corresponds to a transition between different RT surfaces. The transition for the pure case occurs at $l = \pi$, ensuring that the entropy goes to zero as the interval encompasses the entire CFT. The homology constraint is manifestly not imposed, as there is no way to continuously deform the RT surface through the black hole. For the mixed state, the discontinuity occurs at some $l > \pi$ set by the mixing term $\ln M$. Choosing $M = e^{2 \pi^2 c/3 \beta}$, $S_{\mathrm{Mixed}}$ reproduces the entropy of an interval in the thermal state. The transition found here is exactly the bulk RT transition from a single surface into two disconnected surfaces in the eternal black hole geometry. We see that for $l = 2 \pi$, we get the area of the horizon result consistent with the homology constraint. This behavior is displayed pictorially in figure \ref{homologyM}.

Let us consider intermediate values of $\ln M$. For $\ln M \ll c$, the transition occurs at almost $l \sim \pi$ for the mixed case, and so there is almost no difference between the two states. It then seems that the homology constraint is not imposed\footnote{In this case, one can perhaps continue to assume that the homology constraint was imposed but that the circumference of the `wormhole' was too small, $\ll O(c)$ in Planck units, to cause a significant shift in the jump of the RT surface.}. An interesting case to consider is when $\ln M  = {2 \pi^2 c / 3 \beta'} $, with $\beta' > \beta$ and does not depend on $c$. In this case, there will be an appreciable distinction between the two states. After the transition, the mixed state entropy will be
\begin{align}
S_{\mathrm{Mixed}} ={2 \pi^2 c \over 3 \beta'}  + {c \over 3} \ln\left[{\beta \over \pi \epsilon} \sinh \left( {(2 \pi - l) \pi \over \beta}  \right)\right]
\end{align}
The second piece of this expression describes the usual RT surface anchored on the complement of the interval. For $\beta' = \beta$, the first term is the horizon area of the black hole. However, for $\beta' > \beta$, this contribution is smaller than the area, and there is no closed bulk minimal surface outside the black hole that can reproduce it. Naively, this would say that there there is no bulk prescription for such a state, and this `thermal' piece needs to be added in by hand. Moreover, it would seem that the homology constraint is not satisfied.

Incidentally, we know of two sided states which behave very much like $|\mathrm{Mixed} \ran$ for $\beta' > \beta$. These are the Shenker-Stanford wormholes constructed in \cite{Shenker:2013pqa, Shenker:2013yza}. By acting on the TFD with a series of anti-time-ordered shockwaves, they produce a state dual to an elongated wormhole. If the shock waves are sent in symmetrically from the two sides, then the original eternal black hole bifurcate horizon migrates into the wormhole. This region inside the wormhole is known as the causal shadow of the two boundaries. The sizes of the black hole event horizons, those seen by the CFTs, are not representative of the entanglement between them; they only measure how much energy was sent into the wormhole. The bifurcate horizon continues to be the extremal surface and its area correctly measures the entanglement between the CFTs. This is expected since sending in the shock waves amounts to acting on $\cal{H}_L \otimes \cal{H}_R$ with a factorizable unitary which does not modify the entanglement entropy.

\begin{figure}
\begin{center}
\includegraphics[height=3.5cm]{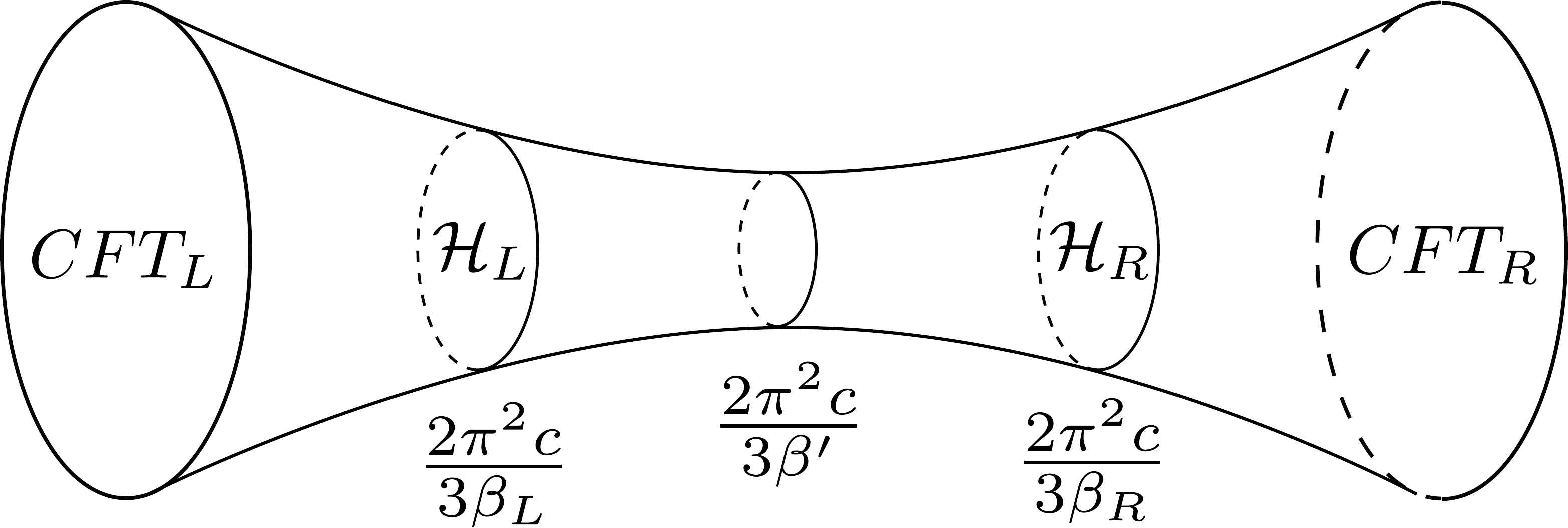}
\caption{The spatial geometry that passes through the bifurcate horizon of an elongated wormhole. $2 \pi^2 c/ 3 \beta'$ represents the entanglement entropy between the two CFTs. This surface is hidden behind both the left and right horizons, and will not be visible in any RT prescription restricted to the exterior of them.}\label{wormhole}
\end{center}
\end{figure}

We could also have considered states which behave like asymmetric wormholes, by considering an asymmetric entangled state with the dimensions of the left and right operators differing by $O(c)$. Note, the maximum number of terms in such a state will be given by the density of states of the side with smaller dimension. From the bulk perspective, the two exterior horizons have different sizes, and again the entanglement is given by the original bifurcate horizon. Figure \ref{wormhole} shows what a spatial slice in this geometry looks like. In the situation where the number of terms saturates the density of states of one side, the entanglement entropy will be the horizon area of that same side. In the Shenker-Stanford construction, this is a state produced by sending in shockwaves from a single side.

In both of these cases, the constant piece in the entropy, $2 \pi^2 c/ 3 \beta'$, plays the role of the area of the original horizon and will not be visible from any of the exteriors. We should stress that the comparison between the state $| \mathrm{Mixed} \ran$ and the Shenker-Stanford wormholes is merely an analogy; the state could instead be dual to a bulk with no geometric description behind the horizons\footnote{We thank Susskind for pointing out that the boost symmetry of the state $|\text{Mixed}\rangle$ makes an interpretation of this state as a long wormhole subtle. The long wormholes of Shenker and Stanford do not have such a symmetry.}. Perhaps, one can get to the Shenker-Stanford states by acting on  $| \mathrm{Mixed} \ran$ with a factorizable unitary on the two sides. One can view this large degree of entanglement between the two CFTs as being large enough to \emph{possibly} describe a geometric connection between the two sides \cite{Maldacena:2013xja}. This is a re-emphasis of the statement that not any entanglement is enough to have a geometry, but a specific kind of one \cite{Almheiri:2012rt, Almheiri:2013hfa}.

\section{Entropy Operators More Generally}
\label{entropy-coarse}

The preceding discussion established that, for 1+1 CFTs dual to three dimensional Einstein gravity, the entanglement entropy of an interval in subspaces of dimension much less than $e^{O(c)}$ could be interpreted as the expectation value of a linear operator acting within that subspace. The approximate linearity of the entropy was established under the assumption of Virasoro identity block dominance. But, as we now discuss, approximate linearity is expected to hold much more generally. It should certainly hold for Einstein gravity in any dimension. In fact, a version of it should hold in any large $N$ theory with many local degrees of freedom.

In the large $N$ limit certain quantum variables become non-fluctuating and a preferred set of ``classical" states is selected. Moreover, the entropy of a subsystem $R$ of a state $|\psi\rangle$ typically becomes large: if $N$ denotes the extensive parameter, then
\beq
\lim_{N \rightarrow \infty} \frac{S_R(|\psi\rangle)}{N} = s_R(|\psi\rangle) > 0 ,
\eeq
in terms of the entropy density $s_R$. In this sense, the main point of large $N$ is that it defines a small parameter, $1/N$, such that the leading contribution to the entropy can be interpreted, within some bounds, as a linear operator.

To illustrate the broad ideas, consider the following general setup. Given a bi-partite system $AB$, we can choose a set of states $D = \{|\psi_1\rangle, ... ,|\psi_K\rangle \}$ and a set of projective measurements on $A$, $M_A = \{ P_1, ..., P_K \}$, and on $B$, $M_B = \{ Q_1, ..., Q_K \}$ such that
\beq \label{distinguish}
P_i |\psi_j \rangle = Q_i |\psi_j\rangle = \delta_{ij} |\psi_j\rangle.
\eeq
That is, the projectors $P_i$ and $Q_i$ serve to distinguish the states in $D$ on both $A$ and $B$. The largest $K$ can be (if we demand perfect distinguishability) is the smaller of the two Hilbert space dimensions of $A$ and $B$. The equivalent statement in the holographic set-up is that states of different entropy can be distinguished using the area operator, i.e. if $\hat{\cal{A}} = \sum_{\cal{A}} \cal{A} |\cal{A}\rangle \langle \cal{A} | $ is a spectral decomposition of the area operator, then we may take the projectors $P = |\cal{A}\rangle \langle \cal{A} |$.

What \eqref{distinguish} says is that the states in $D$ are perfectly distinguished by the measurements in $M_A$ and $M_B$. Furthermore, the measurements are non-destructive or gentle in the sense that the final state after the measurement is the same as the initial state. More importantly, even if the set of states of interest only satisfy \eqref{distinguish} approximately, it can still be true that the large $N$ part of the entropy is correctly reproduced by a linear operator. In the holographic setup, these statements are a reflection of the fact that the area operator becomes non-fluctuating at large $c$. Given this data, as well as the list of entropies $S_A = S_B$ of the $|\psi_i\rangle \in D$, we can form the operators
\beq\label{genentop}
\hat{S}_A = \sum_{i \in D} S_A(|\psi_i\rangle) P_i
\eeq
and
\beq
\hat{S}_B = \sum_{i \in D} S_B(|\psi_i\rangle) Q_i.
\eeq
It follows immediately from \eqref{distinguish} that
\beq
\tr(|\psi_i\rangle \langle \psi_i| \hat{S}_{A,B} ) = S_{A,B}(|\psi_i\rangle).
\eeq

Now suppose we take take a superposition of states in $D$, e.g.
\beq
|\psi\rangle = \alpha | \psi_1 \rangle + \beta |\psi_2 \rangle.
\eeq
Upon tracing out region $B$, the state on subsystem $A$ is
\beq
\psi_A = \text{tr}_B( \alpha^2 |\psi_1 \rangle \langle \psi_1 | + \alpha \beta |\psi_1 \rangle \langle \psi_2 | +  \alpha \beta |\psi_2 \rangle \langle \psi_1 | + \beta^2 |\psi_2 \rangle \langle \psi_2 |).
\eeq
To simplify the form of $\psi_A$ we use the existence of the projective measurements $M_B$. Without changing the value of the trace, we may insert a resolution of the identity on $B$ which contains the projectors $Q_1$ and $Q_2$. Since $\text{tr}(Q_1 |\psi_2 \rangle \langle \psi_1 |) = \text{tr}(Q_2 |\psi_2 \rangle \langle \psi_1 |)= 0 $, it follows that
\beq
\psi_A = \alpha^2 \psi_{A2} + \beta^2 \psi_{A1}.
\eeq
Hence superpositions on the full system reduce to mixtures on a subsystem. This statement is also approximately true given an appropriate approximate form of \eqref{distinguish}.

Consider now a general mixture, $\sigma_A = \sum_i p_i \psi_{Ai}$. Computing $\tr(\hat{S}_A \sigma_A)$ gives
\beq
\tr(\hat{S}_A \sigma_A ) = \sum_i p_i S_A(|\psi_i\rangle).
\eeq
On the other hand, the entropy of $\sigma_A$ is
\beq
S(\sigma) = - \tr(\sigma_A \log \sigma_A).
\eeq
Inserting a resolution of the identity on $A$ that includes the $P_i$, we can write $\sigma = \sum_i p_i \psi_{Ai} = \sum_i p_i P_i \psi_{Ai} P_i$; using \eqref{distinguish} the entropy formula collapses to a single sum
\beq \label{entropymix}
S(\sigma) = - \sum_i \tr( p_i \psi_{Ai} \log(p_i \psi_{Ai})) = \sum_i p_i S_A(|\psi_i \rangle) - \sum_i p_i \log p_i.
\eeq
Provided the second term, called the entropy of mixing and seen earlier in \ref{tfdresult}, is small, the entropy is approximately the average of the entropies of the individual terms.

We will now demonstrate the above logic using three concrete models. Besides illustrating the general discussion, these models will allow us to elucidate the physics of approximate distinguishability.

\subsection{$N$ Copies of a Qubit}
\label{subsec:Nqubits}

As a first toy example, consider states of $N$ qubits of the form
\beq
\rho = \psi^N
\eeq
with
\beq
\psi = \frac{1+\vec{r}\cdot \vec{\sigma}}{2}
\eeq
where $\vec{\sigma} = (\sigma^x,\sigma^y,\sigma^z)$ are Pauli matrices. Such states arise as follows: consider a bipartite system $AB$ where each of $A$ and $B$ consist of $N$ qubits. We restrict attention to states of $AB$ of the form $|\psi\rangle^{ N}$ where $|\psi\rangle$ is an arbitrary pure state of two qubits. Upon tracing out $B$ the resulting state on $A$ is of the form $\rho = \psi^N$.

We will construct a linear operator $\hat{S}$ independent of $\psi$ with the property that
\beq
\tr(\hat{S} \psi^N ) \approx N S(\psi)
\eeq
as $N\rightarrow \infty$. For notational simplicity we will also drop the subsystem index. The idea is that with many copies of the state $\rho$ we can measure $r=|\vec{r}|$ without knowing the eigenvalue basis of $\rho$ and without substantially disturbing the state. In essence, the states $\psi^N$ and $\psi'^N$ are approximately distinguishable for any fixed $\psi \neq \psi'$ in the large $N$ limit.

Think of the $N$ qubits as spin-1/2 operators and introduce the total spin
\beq
J^\alpha = \sum_i \frac{\sigma^\alpha_i}{2}
\eeq
where $\sigma^\alpha_i$ is the $\alpha = x,y,z$ Pauli matrix of qubit $i$. It is straightforward to calculate
\beq
\tr(\psi^N J^\alpha) = \frac{N r^\alpha}{2}
\eeq
and
\beq
\tr(\psi^N J^\alpha J^\alpha ) = \frac{N}{4} + \frac{N(N-1) r^\alpha r^\alpha}{4}.
\eeq
The last line suggests that measuring $J^2 = \sum_\alpha J^\alpha J^\alpha$ effectively measures $r$. Indeed, if the eigenvalues of $J^2$ are $j(j+1)$ then at large $N$ the approximate relation $j = \frac{N r}{2}$ holds. Furthermore, the distribution of $j$ is tightly peaked in the state $\psi^N$ so that $j$ is effectively a semi-classical variable (this can be seen by computing the variance of $|\vec{J}|^2/N^2$).

Let $P_j$ denote the projector onto the eigenspace with $J^2=j(j+1)$ and let
\beq
H(x) = - x \log x - (1-x) \log (1-x)
\eeq
be the Shannon entropy of the probability distribution $\{x,1-x\}$. Then the operator $\hat{S}$ may be taken to be
\beq
\hat{S} = \sum_j N H\left(\frac{1 + 2j/N}{2}\right) P_j
\eeq
where in essence we measure $J^2$ and then return the entropy that would arise for that value of $r = 2j/N$. One easily checks that to leading order in large $N$ we have
\beq
\tr(\hat{S} \psi^N) = N S(\psi).
\eeq

We can also show that the measurement of $\hat{S}$ hardly disturbs the state. Indeed, suppose we measure $\hat{S}$ with relative precision $\epsilon$. Then the question of disturbance amounts to computing
\beq
1-\delta_\epsilon = \tr(P_{|\hat{S} - \langle \hat{S}\rangle| < \epsilon \langle \hat{S}\rangle} \psi^N).
\eeq
The distribution of $j$ in state $\psi^N$ is tightly peaked and Gaussian,
\beq
p(j) \approx \frac{e^{- \frac{(j-\langle j\rangle)^2}{2N \sigma^2}}}{\sqrt{2 \pi N \sigma^2}},
\eeq
where the variance per qubit $\sigma^2$ depends on $r$ but is order one. Converting from $j$ to $S$ (eigenvalue of $\hat{S}$) is accomplished by expanding $H((1+2j/N)/2)$ near $j=Nr/2$. Writing $j = Nr/2 + N y$ we find $S - \langle \hat{S} \rangle = N y H' = (j - Nr/2) H'$, so deviations of the entropy correspond to deviations of $j$ up to a factor of $H'$. Using the distribution for $j$ and the linear change of variables from $j$ to $S$ yields
\beq
\delta_\epsilon \sim \exp(- N k \epsilon^2)
\eeq
for some $r$-dependent constant $k$. For any fixed $\epsilon$ the disturbance caused by the measurement rapidly goes to zero as $N \rightarrow \infty$.

The fact that $\psi^N$ is mostly supported on a projector $P_{|\hat{S} - \langle \hat{S}\rangle| < \epsilon \langle \hat{S}\rangle}$ which is independent of the direction of $\vec{r}$ and hence the basis which diagonalizes $\psi$ has significance. It means that there is a ``universal compression algorithm" \cite{schurweyl1,schurweyl2,universalcomp1,universalcomp2} depending on the spectrum of $\psi$ but not the eigenbasis of $\psi$ which compresses $\psi^N$ into $N S(\psi)$ qubits. In other words, the same compression procedure works for all $\psi$ of the form $\psi = u \psi_0 u^\dagger$ with $u \in \text{SU}(2)$. Explicitly, the algorithm instructs us to make a coarse measure of $j$, after which the state is approximately contained in one of the $j$ eigenspaces of dimension of order $e^{N S(\psi)}$. Furthermore, the probability to obtain a result for $j$ corresponding to a significantly larger than expected dimension is very small. For our purposes, these observations amount to the statement that one doesn't need need to know the basis to measure the entropy.

Hence for this class of states with $N$ taken large there is a linear operator $\hat{S}$ which is independent of the state, semi-classical, and whose expectation value gives the entropy. Furthermore, the linear span of states of the form $|\psi\rangle_{AB}^N$ is the totally symmetric subspace of dimension $N+1$ (recall that each $|\psi\rangle_{AB}$ is a two qubit state). Hence we obtain a polynomially large in $N$ number of states for which an entropy operator exists.

\subsection{Thermal States}

The previous subsection dealt with $N$ copies of a state and showed that in the large $N$ limit an entropy operator existed. However, the copies were strictly non-interacting with each other; in other words, if we view $\psi^N$ as the thermal state of a Hamiltonian, then that Hamiltonian would have no interactions between the qubits. Thus it is useful to give a more general example. There is again a large $N$ limit, a thermodynamic limit, but the ``copies" are no longer non-interacting.

Consider the thermal state of a local Hamiltonian on $N$ qubits in the limit $N \rightarrow \infty$. The state is
\beq
\rho(T) = \frac{e^{-H/T}}{Z(T)}
\eeq
where $Z(T)$ is the partition function. This state describes one half of the thermofield double, and can also be viewed as modeling the coarse grained state of an old black hole. Constructing an entropy operator for this set of states (indexed by temperature) thus improves on the $N$ qubit model described above and shows that the strict independence of the $N$ copies is not required.

Let $S(E,\Delta E)$ denote the microcanonical entropy, the logarithm of the number of energy eigenstates with energy between $E-\Delta E$ and $E+\Delta E$, and let $P_{E,\Delta E}$ denote the projector onto energy eigenstates with energy between $E-\Delta E$ and $E+\Delta E$. Denote by $bins$ a set of such energy windows which completely cover the spectrum of $H$. Finally, define the entropy operator to be
\beq
\hat{S} = \sum_{\text{bins}} S(E,\Delta E) P_{E,\Delta E}.
\eeq

Then we calculate
\beq
\tr(\hat{S} \rho(T)) = \sum_{\text{bins}} S(E,\Delta E) \tr(P_{E,\Delta E} \rho(T)) \approx \sum_{\text{bins}} S(E,\Delta E) \delta_{E,E(T)} \approx S(E(T)) = S(T).
\eeq
where $E(T)$ is the average energy at temperature $T$. As usual, the bin width $\Delta E$ does not play a crucial role; the above identification is valid to leading order in large $N$, with for example $\Delta E = O(1)$. Henceforth we suppress the bin width and distinguish between $S(E)$ and $S(T)$ by context. One can also argue that to leading order in $N$ the measurement of $\hat{S}$ does not disturb the state so that $\hat{S}$ is a semi-classical variable.

Thus there exists a linear operator $\hat{S}$ independent of $T$ which measures the entropy of the family of states $\rho(T)$. It is tempting to identify this operator with the extremal area operator (in Planck units) of the black hole horizon. This identification seems almost trivial, but it is in principle no different from what we did above since $N H((1+2j/N)/2)$ effectively counts the logarithm of the number of states with eigenvalue $j$.

Two important properties of $\hat{S}$ are that it is a coarse-grained observable and that it behaves properly under superposition. We first address the behavior of $\hat{S}$ on mixtures, then we discuss the coarse-grained properties, and finally return to general superpositions.

Consider a mixture of two thermal states,
\beq
\rho = p \rho(T) + (1-p) \rho(T'),
\eeq
and suppose $\rho(T)$ and $\rho(T')$ are distinguishable (this will be true provided $T-T'$ does not vanish too fast as $N\rightarrow \infty$). Then the entropy of $\rho$ is simply
\beq
S(\rho) = p S(\rho(T)) + (1-p) S(\rho(T')) + H(p)
\eeq
where again $H(x)$ is the Shannon entropy defined above. Now compare this to the expectation value of $\hat{S}$ in the state $\rho$. We find
\beq
\tr(\hat{S} \rho) = p S(T) + (1-p) S(T')
\eeq
without the subleading $H(p)$ term. Hence to leading order in the thermodynamic limit, the entropy of the mixture is reproduced by the linear operator. However, this behavior is just what we expect (see Sec. \ref{subsec:areaop_predict},  \ref{subsec:supereternal}) if the entropy operator is proportional to the extremal area operator of the black hole since the expected value of the area is just the weighted average of the two black hole areas.

To show that $\hat{S}$ is a coarse-grained observable we should consider the case where $\rho$ is not a thermal state but is instead a microstate of the thermal ensemble $\rho = |E\rangle \langle E|$. Then we compute
\beq
\tr(\hat{S} |E\rangle \langle E|) = S(E),
\eeq
so in a microstate the entropy operator nevertheless returns the microcanonical entropy, hence it is a coarse-grained observable which returns the coarse-grained entropy not the fine-grained entropy. Furthermore, it is clear that in these cases what $\hat{S}$ is doing is measuring the energy density and returning the appropriate entropy, a manifestly linear operation. That the entropy operator is a coarse-grained observable also matches holographic expectations since the extremal area operator is built from the metric which is in turn constructed from the stress tensor.

Finally, consider a general superposition of the form $\rho = (\sqrt{p} |E\rangle + \sqrt{1-p} |E'\rangle)(\sqrt{p} \langle E| + \sqrt{1-p} \langle E' |)$. The expectation of the entropy operator is
\beq
\tr(\hat{S} \rho) = p S(E) + (1-p) S(E').
\eeq
Similarly, for the mixed state $\rho = p |E\rangle \langle E| + (1-p) |E'\rangle \langle E'|$ the fine grained entropy is $H(p)$ (the binary Shannon entropy) while the expectation of the entropy operator is the same as for the superposition. Note, however, that measurement of the entropy operator collapses the coherent superposition and leaves the mixed state behind if $E$ and $E'$ are in different bins.

From these calculations one sees that the entropy operator can only behave as expected when acting on thermal states, microstates, and mild superpositions or mild mixtures of these. If we begin to make generic superpositions of substantial numbers of microstates then the entropy operator will no longer capture the coarse-grained entropy to leading order in $N$.

\subsection{$N$ Copies of a Free Field Theory}

In this subsection we present one final example of the general construction outlined in the introduction to this section; essentially we study the free limit of a large $N$ vector model.

For simplicity, consider a general bipartite system $AB$ consisting of $k_A + k_B$ free fermion modes with creation and annihilation operators $c_\alpha^\dagger$ and $c_\alpha$ ($\alpha = 1, ... , k_A + k_B$) obeying the algebra $\{c_\alpha , c_\beta^\dagger\} = \delta_{\alpha,\beta}$. Then take $N$ copies of these modes, labelled $c_{\alpha i}$, to give the full algebra $\{c_{\alpha i} , c^\dagger_{\beta j} \} = \delta_{\alpha, \beta} \delta_{i, j}$ defined on the composite system $A^N B^N$. The state of $A^N B^N$ is assumed to be $N$ copies of a single pure Gaussian state of the original $k_A + k_B$ fermion modes.

Upon tracing out subsystem $B^N$, the state on subsystem $A^N$ has the form
\beq
\rho_{A^N} \propto \prod_{i = 1}^N e^{- c^\dagger_i h c_i}
\eeq
where $c^\dagger_i h c_i = c_{\alpha i}^\dagger h_{\alpha \beta} c_{\beta i}$ with the restricted label set $\alpha, \beta = 1, ... , k_A$. The quadratic form of the reduced density matrix is guaranteed due to the initial Gaussian pure state, i.e. because of Wick's theorem.

Now we would like to construct a linear operator that measures the entropy of $\rho_{A^N}$ for any $h$. First, note that if we knew the basis of fermion modes in which $h$ was diagonal, then this problem would be trivial because the problem reduces to decoupled two-level systems, i.e. to qubits. The challenge, as with the $N$ qubit model, is to find a way to measure the spectrum without knowing the basis. To accomplish this measurement, we use a little group theory. To the best of our knowledge our result is new, but we note that similar technology has been used on free bosonic models \cite{marginalprobbosons1,marginalprobbosons2} as part of the ``quantum marginal problem" \cite{marginalprob1,marginalprob2,marginalprob3,marginalprobssa}.

Let $k \equiv k_A$ denote the number of modes in the subsystem. The correlation matrix of a single copy, defined as
\beq
G_{\alpha \beta} = \langle c_\alpha^\dagger c_\beta \rangle_{\rho_A},
\eeq
is a $k\times k$ matrix which is one-to-one with the matrix $h$, $G = (e^{h^T}+1)^{-1}$. The entropy of a single copy can be written in terms of $G$ using the well-known formula
\beq \label{entcorrmatrix}
S_{\text{1 copy}} = - \text{tr}( G \ln G + (1-G) \ln (1-G)).
\eeq
Clearly then if we knew the spectrum of $G$ we could determine the entropy of the $N$ copy system.

However, $G$ itself is not an ideal object to study since it is basis dependent. The basis independent spectrum of $G$ can be obtained from the $k$ numbers $\text{tr}(G^\ell)$ for $\ell =1,...,k$. To construct suitable observables consider the group $U(k)$ of unitary transformations acting on the modes $c_\alpha$. The generators of this group are
\beq
q = c_\alpha^\dagger c_\alpha
\eeq
and
\beq
j^A = c_\alpha^\dagger t^A_{\alpha \beta} c_\beta
\eeq
where $t^A$ are the analogs of the Pauli matrices for $SU(k)$ and $q$ generates the global phase rotation in $U(k)$. Under $U(k)$ transformations $q$ is invariant and $j^A$ transforms in the adjoint representation.

Now on the $N$ copy system we have the corresponding observables
\beq
Q = \frac{1}{N} \sum_i q_i
\eeq
and
\beq
J^A = \frac{1}{N} \sum_i j^A_i.
\eeq
With the factor of $1/N$ these observables are normalized so that their fluctuations vanish in the large $N$ limit. Essentially, this is the generalization of the addition of angular momentum, generalized from $SU(2)$ to $U(k)$. From these observables we construct the $k$ Casimir invariants of $U(k)$,
\beq
C_1 = Q,
\eeq
\beq
C_2 = \sum_A J^A J^A,
\eeq
\beq
C_3 = \sum_{ABC} d_{ABC} J^A J^B J^C,
\eeq
and so on up to the $k$-th Casimir containing $k$ factors of $J^A$\footnote{The $k$-th Casimir may be obtained from the invariant tensor in the $k$-fold tensor product of adjoint representations, i.e. the fusion to the identity of a product of $k$ $J^A$s.}.

The Casimirs, being invariant operators, are not sensitive to the basis which diagonalizes $G$, but they do reveal the spectrum of $G$. For example, if $\lambda_i$ are the eigenvalues of $G$, then the expectation value of $C_1 = Q$ is $\langle C_1 \rangle_N = \sum_i \lambda_i$. The expectation value of a general $C_\ell$ contains terms of the form $\text{tr}(G^n)$ with $n\leq \ell$. Taken together, the expectation values of all the $C_\ell$ suffice to determine the spectrum of $G$. Furthermore, as already mentioned, the fluctuations of the $C_\ell$ vanish in the large $N$ limit, so the spectrum of $G$ becomes in essence a classical variable which can simply be read off from the state without disturbing it.

Since the spectrum determines the single copy entropy via \eqref{entcorrmatrix}, the entropy operator may be taken to be of the form \eqref{genentop} where the projectors are projective measurements of the $k$ Casimir operators constructed above.

To give one simple example of this construction, consider the case $k=2$. Then we are dealing with $U(2)$ and the $t^A$ may be taken to be the Pauli matrices $\sigma^x$, $\sigma^y$, and $\sigma^z$. The correlation matrix $G$ has two eigenvalues, $\lambda_1$ and $\lambda_2$. The expectation values of the Casimirs are
\beq
\langle C_1 \rangle_N = \lambda_1 + \lambda_2
\eeq
and
\beq
\langle C_2 \rangle_N = (\lambda_1 - \lambda_2)^2.
\eeq
Let $P_{c_1,c_2}$ denote the projector onto joint eigenspaces of $C_1$ and $C_2$ labelled by $c_1$ and $c_2$. It is also useful to define the function $\chi(x)$ to be $0$ for $x < 0$, $x$ for $x \in [0,1]$, and $1$ for $x>1$. The binary entropy is again $H(p) = - p \ln p - (1-p)\ln(1-p)$. The entropy operator may then be taken to be
\beq
\hat{S} = \sum_{c_1,c_2} N \left[ H\left(\chi\left(\frac{c_1 + \sqrt{c_2}}{2}\right)\right) + H\left(\chi\left(\frac{c_1 - \sqrt{c_2}}{2}\right)\right)\right] P_{c_1,c_2}.
\eeq

\subsection{Different Sets of States}

As discussed in the homology section and alluded to generally above, one can choose different sets of states to define an entropy operator. For example, one can consider the entropies of subsystems of one side of a two-sided black hole. In this case the homology constraint has an effect because there is a wormhole. We may define an entropy operator which measures the classical geometry outside the black hole and returns the appropriate area in Planck units as the entropy. However, this same entropy operator, when applied to black hole microstates, will still give a entropies appropriate to the corresponding two-sided state. In particular, the homology constraint will not be properly implemented and the entropy of a region and its one-sided complement will not agree.

By the same token, an entropy operator defined for black hole microstates will also not in general function correctly when applied to two-sided states. Of course, this is consistent with everything we said above because these two sets of states are related by superpositions of exponentially many elements. One and two-sided black holes do agree when we restrict to subsystems of less than half the system size. This did not have to be so (it does not follow from just large $N$) but is a consequence of strong coupling (dominance of the identity block). More generally, we would only expect sufficiently small sub-systems to agree between one and two-sided black holes.

Still another interesting class of states is black holes formed by collapse. We can define another linear entropy operator appropriate to these time dependent states, and this operator only sometimes agrees with the operator for two-sided black holes.

\subsection{Recap}

What the results of this section establish is that effective linear entropy operators exist for simple non-interacting large $N$ systems. Moreover, the thermodynamic analysis showed that strictly non-interacting copies were not essential; only something analogous to a thermodynamic limit need exist. The preceding sections established that a linear entropy operator also exists for very strongly interacting large $N$ theories. These data points are suggestive of a more general picture in which the key physics is simply large $N$. Indeed, in the beginning of this section we gave a general argument, framed in terms of gentle distinguishing measurements, that large $N$ was sufficient. The physics is that large $N$ renders appropriate sets of states semi-classical and hence distinguishable. Large $N$ also gives us leave to neglect small entropies of mixing, as in \eqref{entropymix}.

In the case of thermal states indexed by temperature, one could simply measure the energy to gently distinguish different temperatures. For a conformal field theory, this amounts to a measurement of the field theory stress tensor averaged over some region.

For theories that are furthermore holographic and described by Einstein gravity, the stress tensor again plays a privileged role. This is true both for thermal states and more generally. This is because the dual geometry is a natural semi-classical variable that distinguishes different states. Furthermore, in Einstein gravity the geometry is closely related to the field theory stress tensor; a fact reflected in the dominance of the Virasoro identity block in conformal field theories dual to Einstein gravity.

Hence holographic duality has two remarkable aspects: the entropy is a linear operator on certain classes of states (true for all large $N$ theories) and the entropy operator has an incredibly simple interpretation in the dual geometry.

\section{Considerations and Future Directions}

In this paper we have analyzed in some detail the entropy of macroscopic superpositions in semi-classical states within the context of AdS$_3$/CFT$_2$. The main technical tool used was the dominance of the Virasoro identity block in computations of the entropy, a technique that relies on large central charge $c$ and strong coupling (sparse spectrum). We also gave arguments that the same results would be obtained in Einstein gravity in higher dimensions and in fact in a wide variety of systems with an appropriate large $N$ or thermodynamic limit. In this final section we investigate some consequences of our results for certain aspects of quantum information and quantum gravity.

First we note that our extended RT proposal is the same as the recent independent proposal of \cite{2016PhRvD..93h4049P}. They reviewed the standard argument that entropy cannot be a linear operator and argued that the entropy of mild superposition would approximately average assuming at large N that different Schmidt bases were uncorrelated. Our distinguishability arguments include this assumption as a special case and provide a more general information theoretic understanding of entropy as a linear operator. We have also explicitly demonstrated that entropies average for holographic CFT$_2$s and shown how to construct entropy operators for the non-interacting limit of a large N vector model. Thus our analysis includes both weak and strong coupling. Our investigation also considered a number of additional features including the interplay of linearity and homology, the non-linearity of Renyi entropies, and the precise limits of linearity.

\subsection{Conditions for a Semi-Classical Spacetime}

Our results also bear on the entropic approach to bulk reconstruction. For example, it has been found that the leading order in $N^2$ contribution to the tripartite information for any three subregions is nonpositive for any semi-classical holographic state \cite{Papadodimas:2015jra}. However, since the tripartite information is linear in the entropies, the inequality $I_{3} \le 0$ will continue to hold even for superpositions of semi-classical states. In fact, this conclusion holds for the entropy cone of \cite{Bao:2015bfa} since it is closed under averaging.

\subsection{Quantum Error Correction and Superpositions}

Our results imply that we can enlarge the code subspace of states employed in the interpretation that holography is a quantum error correcting code \cite{Almheiri:2014lwa}.  There the code subspace is defined as a space of states perturbatively close to a single reference state (such as the vacuum) that has a semi-classical description in the bulk.  Bulk operators in the entanglement wedge of some boundary region therefore have representations in that region and which act within the code subspace \cite{Dong:2016eik}.  Our results suggest that the code subspace can actually be enlarged to a direct sum of such subspaces, each of which is defined around a different reference state.  Perturbative bulk operators therefore have representations that are block diagonal in the code subspace.

One can prove this last statement provided that the different semi-classical states are distinguishable within the entanglement wedge. Consider a code subspace composed of a direct sum of such distinguishable code subspaces ${\cal{H}}_i$ each defined around a different reference state. Here distinguishable means the states have different geometries and obey \eqref{distinguish} to a high degree of approximation. These code subspaces are \emph{not} perturbatively connected. Next, consider an operator $\phi$ defined in such a way with respect to the boundary that it acts within the entanglement wedge of some region $A$ in all states in the full code subspace. We now show that if the operator $\phi$ satisfies the condition for operator algebra quantum error correction (OAQEC) proved in \cite{Almheiri:2014lwa} for a set of code subspaces ${\cal{H}}_i$ distinguishable within $A$, then it is also satisfied within $\oplus_i {\cal{H}}_i$. In particular, we will show that
\begin{align}
\langle \psi | \big[ \phi, X_{\bar{A}}     \big] |\psi' \rangle = 0
\end{align}
for arbitrary $| \psi \rangle$ and  $| \psi' \rangle$ within $\oplus_i {\cal{H}}_i$, and for any operator $X_{\bar{A}}$ on the complement region $\bar{A}$. We can decompose the states under the direct sum as $|\psi \rangle = \sum_i | c_i \rangle$ and find
\begin{align}
\langle \psi | \big[ \phi, X_{\bar{A}}     \big] |\psi' \rangle = \sum_i \langle c_i |\big[ \phi, X_{\bar{A}}     \big] |c'_i \rangle + \sum_{i \neq j} \langle c_i |\big[ \phi, X_{\bar{A}}     \big] |c'_j \rangle
\end{align}
The first sum vanishes by virtue of $\phi$ satisfying the OAQEC condition within any ${\cal{H}}_i$. We finally argue that the second sum is also zero. Since $\phi$ acts perturbatively within the code subspace the second term is a sum of terms of the form $\langle c_i | X_{\bar{A}} | c'_j \rangle$ where the two states are distinguishable within the region $A$. As discussed earlier around \eqref{distinguish}, this entails the existence of a projection operator purely on $A$ which projects on either of the two code subspaces. Since this operator commutes with any $X_{\bar{A}}$ these matrix elements of $X_{\bar{A}}$ must vanish.

\subsection{A Nonlinearity for Single Sided Pure States}

We demonstrated in section \ref{linhom} that there cannot be a linear entropy operator for all semiclassical states with two asymptotic boundaries. This was primarily due to topology change induced by superposing exponentially many semiclassical states. We argue here that the same obstruction applies in semiclassical states with a single asymptotic boundary.

\begin{figure}
\begin{center}
\includegraphics[height=5cm]{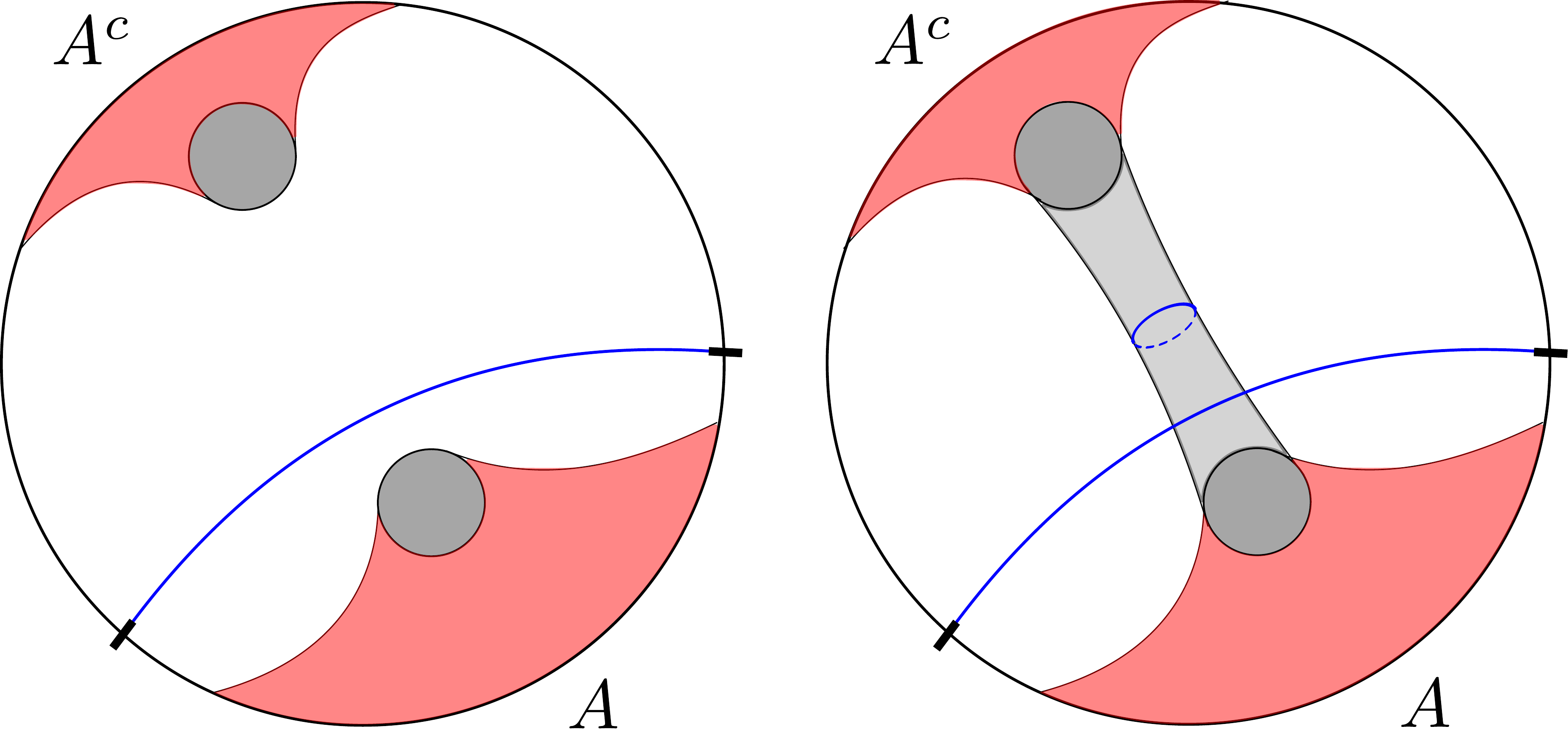}
\caption{Two largely separated black holes in AdS with non-overlapping gravitational dressing. The geometry of the white regions is that of pure AdS. The black holes on the left are two microstates while those on the right are highly entanglement, implying the existence of an Einstein-Rosen bridge.}\label{singlehomology}
\end{center}
\end{figure}

Building on \cite{Susskind:2014yaa}, consider a state that describes two black holes in pure microstates, separated by some large distance in global AdS\footnote{Such a state will not in general be static, but perhaps supersymmetry can be used to obtain one that is such. This point will not affect our argument.}. Moreover, consider the setup where the two black holes have non-overlapping gravitational dressing to two different CFT regions $A$ and $A^c$, as is shown in figure \ref{singlehomology}. Such a state can be created by acting on the vacuum with a product unitary as
\begin{align}
|\psi_i \rangle = U_A^i U_{A^c}^i | 0 \rangle
\end{align}
where $i$ labels the black hole microstate. These unitaries are chosen such that the states $|\psi_i \rangle$ are distinguishable both on $A$ and $A^c$ satisfying
\begin{align}
\text{tr}_{A^c}\left[ U^i_{A^{c}}|0\rangle \langle 0 |  U^{j \dagger}_{A^{c}} \right] = \delta^{i j} \rho_A \label{distingrho}
\end{align}
and similarly for $A$. Since the state is prepared by a product unitary on $A$ and $A^c$, the entanglement entropy of any of those regions will be exactly that of the vacuum. As shown in figure \ref{singlehomology}, the RT surface is simply that of vacuum AdS. Since by the entanglement wedge reconstruction proposal the area operator can be viewed as supported either on $A$ or $A^c$ it will be degenerate within the subspace spanned by  $| \psi_i \rangle$ with its eigenvalue given by that in the vacuum. So we can write
\begin{align}
\hat{S}_A = \sum_i S_A(|0\rangle) | \psi_i \rangle \langle \psi_i | \label{entopsinglehom}
\end{align}
where $S_A(|0\rangle)$ is the entanglement entropy in the vacuum state.

Consider now the superposition of the states $| \psi_i \rangle$ involving all the microstates of the black hole. Here we are restricting to some energy window that involves summing over an exponential number of states. This is
\begin{align}
|w\rangle = {1 \over \sqrt{M}}\sum_i^{M} |\psi_i\rangle = {1 \over \sqrt{M}}\sum_i^M U_A^i U_{A^c}^i | 0 \rangle
\end{align}
where $M$ is the number of microstates. This state is expected to be dual to a wormhole connecting the two black holes in global AdS. This is motivated by ER=EPR \cite{Maldacena:2013xja} ideas and is also supported by explicit constructions involving pair creation of black holes via tunneling \cite{Garfinkle:1990eq, Garfinkle:1993xk}. The trace of the replicated density matrix of $A$ is
\begin{align}
\Tr \rho^n_{A} = \sum_i^M {1 \over M^{n}} \Tr \left( \rho^i_{A}\right)^n
\end{align}
where $\rho^i_A = U^i_A \rho_A U^{i \dagger}_A$ and \ref{distingrho} implies that $\rho^i \rho^j = \delta^{i j} (\rho^i)^2$ as an operator statement. This gives the von Neumann entropy
\begin{align}
S_A(|w\rangle) = S_A(| 0 \rangle) + \ln M.
\end{align}
Had we used \ref{entopsinglehom} to compute the entropy we would completely miss the $\ln M$ contribution coming from area of the wormhole which captures the entanglement between the black holes.

\subsection{Connections to One-Shot Information Theory}

Because our arguments relied on a kind of thermodynamic limit, they are related to recent studies of the so-called one shot information theory of quantum field theories \cite{2015JHEP...06..157C}. Standard many-copy information deals with operational tasks like compression in the limit where the states of interest consist of many independent copies of a single state, the model considered in Sec.~\ref{subsec:Nqubits}. One speaks about compression rates, for example: the resources needed per copy to compress many copies of a state. The resources needed in the single copy limit are typically different, but in many cases the existence of a thermodynamic limit in a single copy setting is sufficient to effectively be in the many copy limit. It would be interesting to further explore these connections as part of the burgeoning one-shot information theory of quantum fields. A concrete question concerns the possibility of universal compression, similar to known results in the $N$ qubit model, but perhaps based on representations of the conformal group instead of the permutation group. One application of these ideas is the justification of the oft made assumption that one may reason about holographic entanglement by simply ``counting Bell pairs".

\subsection{Tensor Networks for Superpositions}

Another interesting direction relates to tensor network models of AdS/CFT. Because of the connection between tensor networks and geometry, it is a prediction of our work that superpositions of macroscopically distinct tensor networks obey the extended RT proposal in its network form. One setting where this prediction can be tested is the random tensor networks models introduced in \cite{constructholospacetimes} and generalized and studied in detail in \cite{randomtensorholo}. Some care must be exercised, since the simplest random tensor network calculations involve not the entanglement entropy but the second Renyi entropy (which does not behave as a linear operator as we show below). However, the general distinguishability arguments should apply to random tensor networks, so we expect that the extended RT proposal does apply to random tensor networks. One simplified setting where this could be explicitly checked consists of so-called random stabilizer tensor networks. Every subsystem density matrix of a stabilizer network has a flat spectrum, so the analysis of random stabilizers is considerably simpler than for generic random tensors. It is also interesting to explore the construction of a more elaborate single tensor network which encodes a superposition of simpler tensor networks.

\subsection{Comment on (Non)linearity of Renyi entropies}

Finally, we briefly comment on the inherent non-linearity of the Renyi entropy. Recall that the Renyi entropy $S_n$ is defined as
\beq
S_n = \frac{1}{1-n}\log \tr(\rho^n).
\eeq
It is usually assumed in field theory calculations that the limit $n\rightarrow 1$, which recovers the von Neumann entropy, is smooth. In fact, the identity block calculations above are only really controlled in the limit $n\rightarrow 1$ with $c(n-1)$ kept large.

Here we show that for superpositions of the type we have been considering the Renyi entropy is badly discontinuous as a function of $n$ if the large $N$ limit is taken first. For simplicity consider two states $\rho_a$ and $\rho_b$ with no overlap,
\beq
\rho_a \rho_b = 0.
\eeq
Further suppose that these states have a flat spectrum with entropies $S(\rho_a) = S_a =N  s_a $ and $S(\rho_b) = S_b = N s_b $.

The Renyi entropy of the state $\sigma = p \rho_a + (1-p)\rho_b$ is
\beq
S_n(\sigma) = \frac{1}{1-n}\log\( p^n e^{-(n-1)S_a} + (1-p)^n e^{-(n-1)S_b}\).
\eeq
To gain intuition set $p=1/2$; the expression inside the logarithm has drastically different behavior depending on whether $n < 1$ or $n>1$. Suppose without loss of generality that $S_a \geq S_b$. Then for $n<1$ the Renyi entropy is
\beq
S_{n<1} = S_a + \frac{n}{1-n} \log 2 + O\( e^{-N(n-1) (s_a - s_b)}\).
\eeq
Hence for fixed $n \neq 1$ the limit $N \rightarrow \infty$ gives $S_{n<1} = S_a$. For $n>1$ the $S_a$ term inside the logarithm is now exponentially smaller than the $S_b$ term. Hence
\beq
S_{n>1} = S_b + \frac{n}{1-n} \log 2 + O\( e^{-N(n-1) (s_b - s_a)}\),
\eeq
and the large $N$ limit again produces a discontinuity.

These results are not an artifact of setting of $p=1/2$. For any $p \in (\epsilon,1-\epsilon)$ with $\epsilon$ fixed, the Renyi entropy is discontinuous as $N \rightarrow \infty$.

\acknowledgments{We are indebted to Joseph Polchinski and his visceral uneasiness with $S = A$ which prompted this work. It is a pleasure to thank Ethan Dyer, Guy Gur-Ari, Daniel Harlow, Raghu Mahajan, Michael Walter, Aitor Lewkowycz, and Lenny Susskind for illuminating discussions. AA emphasizes his gratitude to Ethan Dyer for teaching him how to CFT.  The work of XD was supported in part by the National Science Foundation under Grant No. PHY-1316699, by the Department of Energy under Grant No. DE-SC0009988, and by a Zurich Financial Services Membership at the Institute for Advanced Study. BGS is supported by funds from the ``It From Qubit" Simons Collaboration and from CIFAR.}

\appendix

\section{Entanglement Entropy for a Semi-Classical Superposition}
\label{AppSuper}

In this appendix we present our calculation of the Virasoro identity block contribution to the entanglement entropy of an interval in a state of the form
\begin{align}
\sum_i \alpha_i \mO_i(0) | 0 \rangle.
\end{align}
where the operators $\mO_i$ will be assumed to be primary operators. We comment above in sections \ref{review} \& \ref{onesidedsuper} when we expect the Virasoro identity block contribution to be a good approximation to the entanglement entropy.

As reviewed above in the main text, the entropy is given by
\begin{align}
S = - \partial_n \Tr \rho^n |_{n \rightarrow 1}.
\end{align}
where the density matrix in the replicated manifold is given by
\begin{align}
\Tr \rho^n &=  \langle 0 |\left(\sum_i \alpha_i^* \left(O_i(0)\right)^\dagger\right)^n \sigma_n \sigma_{-n} \left(\sum_i \alpha_i O_i(0)\right)^n     | 0  \rangle \\
&= \sum_{\substack{a_1, ..., a_{M} = 0 \\ b_1, ..., b_{M} = 0}}^n    \alpha_1^{ a_1} ... \alpha_{M}^{ a_{M}} \alpha_1^{* b_1} ... \alpha_{M}^{ * b_{M}}   \langle 0 | (O_1^\dagger)^{b_1} ... (O_{M}^\dagger)^{b_{M}}  \sigma_n \sigma_{-n} O_1^{a_1} ... O_{M}^{a_{M}}  | 0 \rangle
\end{align}
where
\begin{align}
O_1^{a_1} ... O_{M}^{a_{M}} &\equiv \overbrace{O_1  \otimes...\otimes O_1}^{a_1} \otimes \dots \otimes \overbrace{O_M  \otimes...\otimes O_M}^{a_M} + \ \left( {n! \over a_1! ... a_M!} - 1 \right) \ \mathrm{permutations} \\
&\equiv O_{h(\{ a_i \})}
\end{align}
where $h(\{ a_i \}) = \sum_i a_i h_i$, and $h_i$ are the holomorphic dimensions of the operators $O_i$. Focusing only on the Virasoro identity block contribution and performing the OPE expansions in the t-channel it is clear that the only non-zero contributions will come from terms with $a_i = b_i$. The trace then becomes
\begin{align}
\Tr \rho^n = \sum_{a_1, ..., a_{m} = 0}^n  & |\alpha_1|^{2 a_1} ... |\alpha_{M}|^{2 a_{M}} \delta_{n, \sum_j^M a_j}
\langle 0 | O_{h(\{ a_i \})}^\dagger \sigma_n \sigma_{-n} O_{h(\{ a_i \})} | 0 \rangle
\end{align}
Before proceeding we note that the operators $O_{h(\{ a_i \})}$ are sums over primary operators with canonically normalized two point functions, $ 1/x^{h+ \bar{h}}$,  and so will not be canonically normalized themselves. We fix this with the following rescaling
\begin{align}
O_{h(\{ a_i \})} \rightarrow \sqrt{n! \over a_1! ... a_M!} O_{h(\{ a_i \})}.
\end{align}
then the trace becomes
\begin{align}
\Tr \rho^n = \sum_{a_1, ..., a_{M} = 0}^n  &\left( { n! \over a_1! ... a_M!} \right) |\alpha_1|^{2 a_1} ... |\alpha_{M}|^{2 a_{M}}
\langle 0 | O_{h(\{ a_i \})}^\dagger \sigma_n \sigma_{-n} O_{h(\{ a_i \})} | 0 \rangle.
\end{align}
Before taking the derivative with respect to $n$ we need to perform the sum over the $a_i$'s. For $n>1$ the  summand involves four point functions of heavy operators  whose Virasoro identity block contribution is not known in closed form. To get around this, we first consider a modified form of the above equation
\begin{align}
\sum_{a_1, ..., a_{M} = 0}^m  &\left( { m! \over a_1! ... a_M!} \right) |\alpha_1|^{2 a_1} ... |\alpha_{M}|^{2 a_{M}}
\langle 0 | O_{h(\{ a_i \})}^\dagger \sigma_n \sigma_{-n} O_{h(\{ a_i \})} | 0 \rangle.
\end{align}
where we replaced $n$ in the upper limit of the sum over $a_i$ and in the combinatoric factor with a new variable $m$. We will tune $m$ and $n$ separately in the meantime and then take the $m \rightarrow n$ limit before differentiating. Next, we take $n$ close to 1 and use the known closed form expression of the identity block for this four point function. These are
\begin{align}
\langle 0 | O_{h(\{ a_i \})}^\dagger \sigma_n \sigma_{-n} O_{h(\{ a_i \})} | 0 \rangle = \left( \sqrt{1 - 24 \sum_i h_i a_i / c n} \over 2 \sin {l \over 2} \sqrt{1 - 24 \sum_i h_i a_i / c n}   \right)^{c (n-1)\over 3} \equiv f(\{a_i \},n,l).
\end{align}
where $l$ is the size of the interval. Note, that here we have specialized to the case of operators with no spin. This function is unfortunately sufficiently complicated that we cannot perform the sum directly. Instead, we perform a Taylor expansion of the function in the size of the interval, $l$, and then preform the sum over $a_i$ term by term. We will see that the Taylor expansion in $l$ is resummable even after differentiating w.r.t. $n$.

Let us first make the following definition
\begin{align}
f(\{a_i \},n,l) = \left( \sqrt{1 - 24 \sum_i h_i a_i / c n} \over 2 \sin {l \over 2} \sqrt{1 - 24 \sum_i h_i a_i / c n}   \right)^{c (n-1)\over 3}  \equiv \left( {g(\{ a_i \}, n, l) \over l} \right)^{c (n-1)\over 3}
\end{align}
The function $g(\{ a_i \}, n, l)$ goes to $1$ as $l \rightarrow 0$. Expanding in $l$ we have
\begin{align}
f(\{a_i \},n,l) = l^{c(1 - n) \over 3} \sum_{k = 0}^\infty {\partial_l^k g^{c (n-1)\over 3}|_{l\rightarrow 0} \over k! } l^k
\end{align}
For the $k$-th derivative of $g^{c (n-1)\over 3}$ we use the formula,
\begin{align}
\partial_l^k g^{c (n-1)\over 3} =k! \sum_{ \{ c_i \} = 0}^k  \left( { \left({c (n-1)\over 3}\right)! \over (c_1)! (c_2)! \dots (c_k)! \left({c (n-1)\over 3} - \sum_i c_i\right)! } \right)  { g^{{c (n-1)\over 3} - \sum_i c_i}  \over \prod_{j=1}^k (j!)^{c_j}}(\partial_l^1 g)^{c_1} (\partial_l^2 g)^{c_2} \dots (\partial_l^k g)^{c_k} \label{taylor}
\end{align}
along with the condition that $\sum_{j = 1}^{k} j \times c_j = k$. The expansion of $g(a,n,l)$ and its derivatives in $l$ are
\begin{align}
\partial_l^k g(\{a_i \},n,l) &= \sum_{p = 0}^\infty {(-1)^{p+1} 2 (2^{2p - 1} - 1) B_{2 p} \over (2 p - k)!} \left( x \over 2 \right)^{2 p} l^{2 p - k}
\end{align}
Where $x = \sqrt{1 - {24 \sum_i h_i a_i \over c n}}$, and $B_{2 p}$ are the Bernoulli numbers. Taking the limit as $l \rightarrow 0$ we get
\begin{align}
\partial_l^k g(\{a_i \},n,l)|_{l \rightarrow 0} &=  {(-1)^{{k \over 2}+1} 2 (2^{k - 1} - 1) B_{k} } \left( x \over 2 \right)^{k} \  \ \ \mathrm{for \ even \  k }\label{even}\\
&= 0\   \ \ \ \ \ \ \ \ \ \ \ \ \ \ \ \ \  \ \ \ \ \ \ \ \ \ \ \  \ \ \ \ \ \ \ \ \ \  \mathrm{for \ odd \ k.} \label{odd}
\end{align}
Plugging this back into \ref{taylor} we get
\begin{align}
\partial_l^k g^{c (n-1)\over 3} &=k! \sum_{ \{ c_i \}_{even} = 0}^k  \left( { \left({c (n-1)\over 3}\right)! \over  (c_2)! (c_4)!\dots (c_k)! \left({c (n-1)\over 3} - \sum_i c_i\right)! } \right)  {1 \over \prod_{j=even}^k (j!)^{c_j}} \times \nonumber \\
& \ \ \ \  \left( (-1) 2 (2^{-1} - 1) B_0 \right)^{{c (n-1)\over 3} - \sum_i c_i} \prod_{i = even}^{k} \left( {(-1)^{i/2 + 1} 2 (2^{i - 1} - 1) B_i \over 2^i} \right)^{c_i} x^k \\
&\equiv k! G_k(n) x^k \label{Gn}
\end{align}
The expansion of $f(\{a_i \},n,l)$  now simplifies to
\begin{align}
 f(\{a_i \},n,l) = \sum_{k = even}^\infty G_k(n) \left(\sqrt{1 - {24 \sum_i h_i a_i \ \over n c}}\right)^k l^{k-{c (n-1)\over 3}}
\end{align}
where the sum over $k$ runs only over the evens because of  $\sum_{j = 1}^{k} j \times c_j = k$ and equations \ref{even} \& \ref{odd} ; since all the $c_{odd}$ terms vanish, $k$ must be even.

Plugging this into the formula for the entropy, we have that
\begin{align}
S &= - \lim_{n \rightarrow 1} \partial_n \lim_{m\rightarrow n}\left(    \sum_{k = even}^\infty G_k(n)  l^{k-{c (n-1)\over 3}}   \sum_{a_1, ..., a_{M} = 0}^m   \left( { m! \over a_1! ... a_M!} \right) |\alpha_1|^{2 a_1} ... |\alpha_{M}|^{2 a_{M}}  \left(\sqrt{1 - {24  \sum_i h_i a_i \over n c}}\right)^k   \right) \label{trrhonexp}\\
&=  -  \sum_{k = even}^\infty \lim_{n \rightarrow 1} \partial_n \lim_{m\rightarrow n} \left(    G_k(n)  l^{k-{c (n-1)\over 3}}   \sum_{a_1, ..., a_{M} = 0}^m   \left( { m! \over a_1! ... a_M!} \right) |\alpha_1|^{2 a_1} ... |\alpha_{M}|^{2 a_{M}}  \left(\sqrt{1 - {24  \sum_i h_i a_i \over n c}}\right)^k   \right)
\end{align}
Thus, we can perform the differentiation and continuation in $n$ before summing over $k$. Let us consider the terms with $k = 0$ and $k \neq 0$ separately.

Before differentiation, the $k = 0$ term is
\begin{align}
G_0(n)  l^{-{c (n-1)\over 3}}   &\sum_{a_1, ..., a_{M} = 0}^m   \left( { m! \over a_1! ... a_m!} \right) |\alpha_1|^{2 a_1} ... |\alpha_{m}|^{2 a_{m}} = G_0(n)  l^{-{c (n-1)\over 3}}  \left( \sum_{i = 1}^M |\alpha_i|^2 \right)^m \\
&= \left( {-2 (2^{-1} - 1)B_0 \over l} \right)^{{c (n-1)\over 3}} \left( \sum_{i = 1}^M  |\alpha_i|^2 \right)^m
\end{align}
The contribution this gives to the entanglement entropy is
\begin{align}
S_0 = \sum_{i = 1}^M  |\alpha_i|^2   \left( {c \over 3} \ln \left(  {l \over -2 (2^{-1} - 1)B_0 }  \right) +  {c \over 3} \ln {1 \over \epsilon} \right)
\end{align}
where we have put back the UV cut-off $\epsilon$. Note here that $\sum_{i = 1}^M  |\alpha_i|^2  = 1$ but we chose to keep it explicit for later purposes. We see that this contribution is simply the average of those of the branches of the wavefunction. This contribution is also the leading part of the entanglement entropy.

Now let us focus on the $k \neq 0$ contributions to the entropy. There are many terms on which $\partial_n$ can act, but note that $G_k(n)$ actually vanishes as $n \rightarrow 1$ making it the only relevant term to act the derivative on. We can further pinpoint exactly which part of $G_k(n)$ the derivative needs to hit. It turns out that the important factor is the $\left( {c (n- 1) \over 3} - \sum_i c_i  \right)!$ in the denominator. When taking $n$ to $1$ this just becomes the factorial of a negative integer, thus blowing up and causing the entire expression to vanish. Taking its derivative we find
\begin{align}
\lim_{n \rightarrow 1}\partial_n {1 \over \left( {c (n- 1) \over 3} - \sum_i c_i  \right)! } = -{c \over 3}\frac{ \psi ^{(0)}(1-\sum_i c_i)}{ \Gamma (1-\sum_i c_i)} = {c \over 3} (-1)^{\sum_i c_i-1} \left(\sum_i c_i-1\right)!
\end{align}
As required, this does not vanish. Taking $m \rightarrow n$ and then the $n \rightarrow 1$ of the other factors in the $k$-th contribution of $S$ we find
\begin{align}
S_k &=- {c \over 3}\left( \lim_{n \rightarrow 1}  \sum_{a_1, ..., a_{M} = 0}^n   \left( { n! \over a_1! ... a_M!} \right) |\alpha_1|^{2 a_1} ... |\alpha_{M}|^{2 a_{M}}  \left(\sqrt{1 - {24  \sum_i h_i a_i \over n c}}\right)^k  \right) l^{k} \times \nonumber \\
& \sum_{ \{ c_i \}_{even} = 0}^k   { (-1)^{\sum_i c_i-1} \left(\sum_i c_i-1\right)! \over  (c_2)! (c_4)!\dots (c_k)!  }   {1 \over \prod_{j=even}^k (j!)^{c_j}} \times \nonumber \\
& \ \ \ \  \left( (-1) 2 (2^{-1} - 1) B_0 \right)^{ - \sum_i c_i} \prod_{i = even}^{k} \left( {(-1)^{i/2 + 1} 2 (2^{i - 1} - 1) B_i \over 2^i} \right)^{c_i}
\end{align}
We still have the first factor we need to evaluate. This sum simply turns out to be
\begin{align}
\sum_{a_1, ..., a_{M} = 0}^n   \left( { n! \over a_1! ... a_M!} \right) |\alpha_1|^{2 a_1} ... |\alpha_{M}|^{2 a_{M}}  \left(\sqrt{1 - {24  \sum_i h_i a_i \over n c}}\right)^k   = \left( \sum_{i = 1}^M |\alpha_i|^2 \left( \sqrt{1 - {24   h_i \over n c}} \right)^k \right)^n \label{suma}
\end{align}
and the $n \rightarrow 1$ limit of which is obvious.

We're almost done now. We found that all the contributions to the entropy separate as contributions from the different branches of the wavefunction. What is left is to see is that the sum over $k$ can actually be done and gives the answer claimed. Focusing on a single branch we have
\begin{align}
S_{i} &=    \left( {c \over 3} \ln \left(  {l \over -2 (2^{-1} - 1)B_0 }  \right) +  {c \over 3} \ln {1 \over \epsilon} \right)  - \left( {c \over 3} \right) \sum_{k = even}^\infty  \sum_{ \{ c_i \}_{even} = 0}^k   { (-1)^{\sum_i c_i-1} \left(\sum_i c_i-1\right)! \over  (c_2)! (c_4)!\dots (c_k)!  }   {1 \over \prod_{j=even}^k (j!)^{c_j}} \times \nonumber \\
& \ \ \ \  \left( (-1) 2 (2^{-1} - 1) B_0 \right)^{ - \sum_i c_i} \left( \sqrt{1 - {24 h_i \over c}} \right)^k \prod_{i = even}^{k} \left( {(-1)^{i/2 + 1} 2 (2^{i - 1} - 1) B_i \over 2^i} \right)^{c_i}
\end{align}
The question now is whether this is expansion resums to anything that we know of. Indeed it does and it resums to
\begin{align}
S_{i} = {c \over 3} \ln\left( {2 \over \epsilon \sqrt{ {24 h_i \over c} - 1}} \sinh\left( {l \over 2} \sqrt{{24 h_i \over c} - 1} \right) \right)
\end{align}
exactly. Thus, the total entropy is
\begin{align}
S &= \sum_{i = 1}^M |\alpha_i|^2 {c \over 3} \ln\left( {2 \over \epsilon \sqrt{ {24 h_i \over c} - 1}} \sinh\left( {l \over 2} \sqrt{ {24 h_i \over c} - 1} \right) \right) \\
&= \sum_{i = 1}^M |\alpha_i|^2 S_i
\end{align}
And the entropy simply averages!

Finally, we perform the check that our method of expanding and resuming preserves the requirement that $\Tr \rho \rightarrow 1$ as $n\rightarrow 1$. From equation \ref{trrhonexp}, we can read off the form of the reduced density matrix
\begin{align}
\Tr \rho^n =    \sum_{k = even}^\infty G_k(n)  l^{k-{c (n-1)\over 3}}   \sum_{a_1, ..., a_{M} = 0}^m   \left( { m! \over a_1! ... a_M!} \right) |\alpha_1|^{2 a_1} ... |\alpha_{M}|^{2 a_{M}}  \left(\sqrt{1 - {24  \sum_i h_i a_i \over n c}}\right)^k
\end{align}
Taking $n \rightarrow 1$ this becomes
\begin{align}
\Tr \rho = \sum_{i = 1}^M | \alpha_i|^2 \sum_{k = even}^{\infty} \left(\lim_{n \rightarrow 1} G_k(n) \right) l^k \left( \sqrt{1 - {24 h_i \over c}} \right)^k
\end{align}
We need to know what is the limit of $G_k(n)$. From equation \ref{Gn} we have
\begin{align}
G_k(n) = {\partial_l^k g^{c (n-1) \over 3} \over k! x^k}
\end{align}
implying that $\lim_{n\rightarrow1} G_k(n) = \delta_{k,0}$. Therefore
\begin{align}
\Tr \rho = \sum_{i = 1}^M |\alpha_i|^2 = 1
\end{align}

\bibliographystyle{jhep}
\bibliography{bibliography}
\end{document}